\documentclass[11pt]{article}

\usepackage{url}

\usepackage{amsmath}
\usepackage{amsthm}
\usepackage{amssymb}

\theoremstyle{plain}
\newtheorem{theorem}{Theorem}[section]
\newtheorem{lemma}[theorem]{Lemma}

\newtheorem{proposition}[theorem]{Proposition}
\newtheorem{definition}[theorem]{Definition}
\newtheorem{remark}[theorem]{Remark}
\newtheorem{example}[theorem]{Example}
\newtheorem{axiom}[theorem]{Axiom}

\newcommand{\abs}[1]{\left\lvert#1\right\rvert}
\newcommand{\ip}[2]{\left\langle#1,#2\right\rangle}
\DeclareMathOperator{\Sim}{sim}

\DeclareMathOperator{\Her}{Her}
\DeclareMathOperator{\HerQ}{Her_{\mathbb{Q}}}

\DeclareMathOperator{\diag}{diag}
\DeclareMathOperator{\rank}{rank}

\newcommand{\N}{\mathbb{N}}
\newcommand{\Q}{\mathbb{Q}}
\newcommand{\R}{\mathbb{R}}
\newcommand{\C}{\mathbb{C}}
\newcommand{\CQ}{\mathbb{C}_Q}
\newcommand{\X}{\Sigma^*}
\newcommand{\K}{H}

\newcommand{\OH}{\hat{H}}
\newcommand{\BO}{\mathcal{B}(X)}
\newcommand{\HO}{\mathcal{B}_h(X)}
\newcommand{\PO}{\mathcal{B}(X)_+}

\newcommand{\fmi}{\varphi}

\newcommand{\noi}{\noindent}

\title{\textbf{
An extension of Chaitin's halting probability $\Omega$\\
to a measurement operator in an infinite dimensional quantum system%
\thanks{
An extended abstract appeared
in the Proceedings of the 6th Conference on Real Numbers and Computers
(RNC'6),
Schlo\ss{} Dagstuhl, Germany, November 15--17, 2004, pp.~172--191.
}
}}

\author{
Kohtaro Tadaki\\
\\
21st Century Center Of Excellence Program, Chuo University,\\
1-13-27 Kasuga, Bunkyo-ku, Tokyo 112-8551, Japan\\
E-mail: tadaki@kc.chuo-u.ac.jp
}

\date{}

\begin{document}

\maketitle

\begin{quotation}
\noi\textbf{Abstract.}
This paper proposes an extension of Chaitin's halting probability $\Omega$
to a measurement operator in an infinite dimensional quantum system.
Chaitin's $\Omega$ is defined as the probability that
the universal self-delimiting Turing machine $U$ halts,
and plays a central role in the development of algorithmic information theory.
In the theory,
there are two equivalent ways to define the program-size complexity $\K(s)$
of a given finite binary string $s$.
In the standard way,
$\K(s)$ is defined as the length of the shortest input string
for $U$ to output $s$.
In the other way,
the so-called universal probability $m$ is introduced first,
and then $\K(s)$ is defined as $-\log_2 m(s)$
without reference to the concept of program-size.

Mathematically,
the statistics of outcomes in a quantum measurement are described by
a positive operator-valued measure (POVM) in the most general setting.
Based on the theory of computability structures on a Banach space
developed by Pour-El and Richards,
we extend the universal probability to an analogue of POVM
in an infinite dimensional quantum system, called a universal semi-POVM.
We also give another characterization of Chaitin's $\Omega$ numbers
by universal probabilities.
Then, based on this characterization,
we propose to define an extension of $\Omega$
as a sum of the POVM elements of a universal semi-POVM.
The validity of this definition is discussed.

In what follows,
we introduce an operator version $\OH(s)$ of $\K(s)$
in a Hilbert space of infinite dimension
using a universal semi-POVM,
and study its properties.
\end{quotation}

\vspace{1mm}

\begin{quotation}
\noi\textbf{Key words:}
algorithmic information theory,
Chaitin's $\Omega$,
quantum measurement,
computable analysis,
POVM,
universal probability
\end{quotation}

\begin{quotation}
\noi\textbf{MSC (2000)}\;
03F60, 68Q30, 81P15, 03D80, 47S30
\end{quotation}

\section{Introduction}

Algorithmic information theory is a framework to apply
information-theoretic and probabilistic ideas to recursive function theory.
One of the primary concepts of algorithmic information theory
is the \textit{program-size complexity}
(or \textit{Kolmogorov complexity}) $\K(s)$ of a finite binary string $s$,
which is defined as the length of the shortest binary input
for the universal self-delimiting Turing machine to output $s$.
By the definition,
$\K(s)$ can be thought of as the information content of
the individual finite binary string $s$.
In fact,
algorithmic information theory has precisely the formal properties of
classical information theory (see \cite{C75}).
The concept of program-size complexity plays a crucial role in
characterizing the randomness of a finite or infinite binary string.
In \cite{C75} Chaitin introduced the halting probability $\Omega$
as an example of random infinite string.
His $\Omega$ is defined as the probability that
the universal self-delimiting Turing machine halts,
and plays a central role in the development of algorithmic information theory.
The first $n$ bits of the base-two expansion of $\Omega$ solves
the halting problem for a program of size not greater than $n$.
By this property,
the base-two expansion of $\Omega$ is shown to be an instance of
a random infinite binary string.
In \cite{C87a} Chaitin encoded this random property of $\Omega$
onto an exponential Diophantine equation in the manner that
a certain property of the set of the solutions of the equation is
indistinguishable from coin tosses.
Moreover, based on this random property of the equation,
Chaitin derived several quantitative versions of
G\"odel's incompleteness theorems.

In \cite{T02}
we generalized Chaitin's halting probability $\Omega$ to $\Omega^D$
so that the degree of randomness of $\Omega^D$ can be controlled by
a real number $D$ with $0<D\le 1$.
As $D$ becomes larger, the degree of randomness of $\Omega^D$ increases.
When $D=1$,
$\Omega^D$ becomes a random real number, i.e., $\Omega^1=\Omega$.
The properties of $\Omega^D$ and its relations to self-similar sets
were studied in \cite{T02}.
In the present paper, however,
we generalize Chaitin's $\Omega$ to a different direction from \cite{T02}.
The aim of the present paper is
to extend Chaitin's halting probability $\Omega$
to a measurement operator in an infinite dimensional quantum system
(i.e., a quantum system whose state space has infinite dimension).

The program-size complexity $\K(s)$ is originally defined
using the concept of program-size, as stated above.
However,
it is possible to define $\K(s)$ without referring to such a concept,
i.e.,
we first introduce a \textit{universal probability} $m$,
and then define $\K(s)$ as $-\log_2 m(s)$.
A universal probability is defined through the following two definitions
\cite{ZL70}.
We denote by $\X$ the set of finite binary strings,
by $\N^+$ the set of positive integers, and
by $\Q$ the set of rational numbers.

\begin{definition}\label{lcsp}
  For any $r\colon \X\to[0,1]$,
  we say that $r$ is a lower-computable semi-measure if
  $r$ satisfies the following two conditions:
  \begin{enumerate}
    \item $\sum_{s\in \X}r(s)\le 1$.
    \item There exists a total recursive function
    $f\colon\N^+\times \X\to\Q$
      such that, for each $s\in \X$,
      $\lim_{n\to\infty} f(n,s)=r(s)$ and
      $\forall\,n\in\N^+\;\>0\le f(n,s)\le f(n+1,s)$.
  \end{enumerate}
\end{definition}

\begin{definition}
  Let $m$ be a lower-computable semi-measure.
  We say that $m$ is a universal probability if
  for any lower-computable semi-measure $r$,
  there exists a real number $c>0$ such that,
  for all $s\in \X$, $c\,r(s)\le m(s)$.
\end{definition}

In this paper we show
that Chaitin's $\Omega$ can be defined using a universal probability
without reference to the universal self-delimiting Turing machine,
as in the case of $\K(s)$.

In quantum mechanics,
a \textit{positive operator-valued measure} (POVM) is the mathematical tool
which describes the statistics of outcomes in a quantum measurement
in the most general setting.
In this paper we extend the universal probability to an analogue of a POVM
in an infinite dimensional quantum system,
called a \textit{universal semi-POVM}.
Then, based on a universal semi-POVM,
we introduce the extension $\hat{\Omega}$ of Chaitin's $\Omega$
to a measurement operator in an infinite dimensional quantum system.

\subsection{Quantum measurements}

Let $X$ be a separable complex Hilbert space.
We assume that the inner product $\ip{u}{v}$ of $X$ is
linear in the first variable $u$ and
conjugate linear in the second variable $v$,
and it is related to the norm by $\|u\|=\ip{u}{u}^{1/2}$.
$\BO$ is the set of \textit{bounded} operators in $X$.
We denote the \textit{identity operator} in $X$ by $I$.
For each $T\in\BO$,
the \textit{adjoint} operator of $T$ is denoted as $T^*\in\BO$.
We say $T\in\BO$ is \textit{Hermitian} if $T=T^*$.
$\HO$ is the set of Hermitian operators in $X$.
We say $T\in\BO$ is \textit{positive} if
$\ip{Tx}{x}\ge 0$ for all $x\in X$.
$\PO$ is the set of positive operators in $X$.
For each $S, T\in\HO$,
we write $S\leqslant T$ if $T-S$ is positive.
Let $\{A_n\}$ be a sequence of operators in $\BO$, and let $A\in\BO$.
We say $\{A_n\}$ \textit{converges strongly} to $A$ as $n\to\infty$ if
$\lim_{n\to\infty}\|A_nx-Ax\|=0$ for all $x\in X$.

With every quantum system
there is associated a separable complex Hilbert space $X$.
The states of the system are described by the nonzero elements in $X$.
In the present paper, we consider
the case where $X$ is a Hilbert space of infinite dimension.
That is, we consider \textit{infinite dimensional quantum systems}.

Let us consider a quantum measurement performed upon a quantum system.
We first define a \textit{POVM on a $\sigma$-field} as follows.
\begin{definition}[POVM on a $\sigma$-field]\label{def-povm-infinie-general}
  Let $\mathcal{F}$ be a $\sigma$-field in a set $\Phi$.
  We say $M\colon\mathcal{F}\to\PO$ is a POVM
  on the $\sigma$-field $\mathcal{F}$
  if the following holds for $M$:
  If $\{B_j\}$ is a countable partition of $\Phi$
  into pairwise disjoint subsets in $\mathcal{F}$,
  then $\sum_j M(B_j)=I$ where the series converges strongly.%
  \footnote{
    In Definition \ref{def-povm-infinie-general} and
    the subsequent Definition \ref{def-povm-countable} and \ref{def-semi-povm},
    we can equivalently replace
    the condition ``the series converges strongly''
    by ``the series converges weakly'',
    using Lemma \ref{bounded monotonic sequence} given below.
    Here,
    for any sequence $\{A_n\}$ of operators in $\BO$ and any $A\in\BO$,
    we say $\{A_n\}$ \textit{converges weakly} to $A$ as $n\to\infty$ if
    $\lim_{n\to\infty}\ip{A_nx}{y}=\ip{Ax}{y}$ for all $x,y\in X$.
  }
\end{definition}
In the most general setting,
the statistics of outcomes in a quantum measurement are
described by a POVM $M$ on a $\sigma$-field in a set $\Phi$.
The set $\Phi$ consists of all outcomes possible under the quantum measurement.
If the state of the quantum system is described
by an $x\in X$ with $\|x\|=1$ immediately before the measurement,
then the probability distribution of the measurement outcomes is given by
$\ip{M(B)x}{x}$.
(See e.g.~\cite{H01}
for the treatment of the mathematical foundation of quantum mechanics.)

In this paper,
we relate an argument $s$ of a universal probability $m(s)$
to an individual outcome which may occur in a quantum measurement.
Thus,
since $m(s)$ is defined for all finite binary strings $s$,
we focus our thought on
a POVM measurement with countably infinite measurement outcomes,
such as the measurement of energy level of a harmonic oscillator.
Since $\Phi$ is a countably infinite set for our purpose,
we particularly define the notion of
a \textit{POVM on a countably infinite set}
as follows.

\begin{definition}[POVM on a countably infinite set]\label{def-povm-countable}
  Let $S$ be a countably infinite set, and let $R\colon S\to\PO$.
  We say $R$ is a POVM on the countably infinite set $S$
  if $R$ satisfies $\sum_{v\in S}R(v)=I$
  where the series converges strongly.
\end{definition}

Let $S$ be a countably infinite set,
and let $\mathcal{F}$ be the set of all subsets of $S$.
Assume that $R\colon S\to\PO$ is a POVM on the countably infinite set $S$
in Definition \ref{def-povm-countable}.
Then, by setting $M(B)=\sum_{v\in B}R(v)$ for every $B\in\mathcal{F}$,
we can show that
$M\colon\mathcal{F}\to\BO$ is a POVM on the $\sigma$-field $\mathcal{F}$
in Definition \ref{def-povm-infinie-general}.
Thus Definition \ref{def-povm-countable} is sufficient for our purpose.
Consider the quantum measurement described by the $R$
performed upon a quantum system.
We then see that
if the state of the quantum system is described by an $x\in X$ with $\|x\|=1$
immediately before the measurement then, for each $v\in S$,
the probability that the result $v$ occurs is given by $\ip{R(v)x}{x}$.
Each operator $R(v)\in\PO$ is called a \textit{POVM element}
associated with the measurement.

In a POVM measurement with countably infinite measurement outcomes,
we represent each measurement outcome by just a finite binary string
in perfect register with the argument of a universal probability.
Thus we consider the notion of a \textit{POVM on $\X$}
which is a special case of a POVM on a countably infinite set.

\begin{definition}[POVM on $\X$]\label{def-povm-infinie}
  We say $R\colon\X\to\PO$ is a POVM on $\X$
  if $R$ is a POVM on the countably infinite set $\X$.
\end{definition}

In a quantum measurement described by a POVM on $\X$,
an experimenter gets a finite binary string as a measurement outcome.

Any universal probability $m$ satisfies $\sum_{s\in \X}m(s)<1$.
This relation is incompatible with the relation $\sum_{s\in\X}R(s)=I$
satisfied by a POVM $R$ on $\X$.
Hence we further introduce the notion of a \textit{semi-POVM on $\X$},
which is appropriate for an extension of universal probability.

\begin{definition}[semi-POVM on $\X$]\label{def-semi-povm}
  We say $R\colon\X\to\PO$ is a semi-POVM on $\X$
  if $R$ satisfies $\sum_{s\in\X}R(s)\leqslant I$
  where the series converges strongly.
\end{definition}

Obviously, any POVM on $\X$ is a semi-POVM on $\X$.
Let $R$ be a semi-POVM on $\X$.
It is easy to convert $R$ into a POVM on a countably infinite set
by appending an appropriate positive operator to $R$ as follows.
We fix any one object $w$ which is not in $\X$.
Let $\widetilde{\Omega}_R=\sum_{s\in\X}R(s)$.
Then $0\leqslant\widetilde{\Omega}_R\leqslant I$
and $\sum_{s\in\X}R(s)+(I-\widetilde{\Omega}_R)=I$.
Thus, by setting $\overline{R}(s)=R(s)$ for every $s\in\X$
and $\overline{R}(w)=I-\widetilde{\Omega}_R$,
we see that
$\overline{R}\colon\X\cup\{w\}\to\PO$ is
a POVM on the countably infinite set $\X\cup\{w\}$
in Definition \ref{def-povm-countable}.
Therefore a semi-POVM on $\X$ has a physical meaning
in the same way as a POVM on a countably infinite set.
Hence, hereafter, we say that a POVM measurement $\mathcal{M}$ is
\textit{described} by a semi-POVM $R$ on $\X$
if $\mathcal{M}$ is described by the POVM $\overline{R}$ on
the countably infinite set $\X\cup\{w\}$.
Let us consider the quantum measurement described by the $R$
performed upon a quantum system.
We then see that
if the state of the quantum system is described by an $x\in X$ with $\|x\|=1$
immediately before the measurement then, for each $s\in\X$,
the probability that the result $s$ occurs is given by $\ip{R(s)x}{x}$.

\subsection{Related works}

There are precedent works which make an attempt to extend
the universal probability to operators in quantum system \cite{G01,T03}.

As we stated above,
in quantum mechanics
a POVM is the mathematical notion
which describes the statistics of outcomes in a quantum measurement
in the most general setting.
Especially in quantum information processing
such as quantum computation, quantum cryptography,
and quantum teleportation and communication
(see e.g.~\cite{NC00} for these subjects),
prior to a real experiment
we design an appropriate POVM
in order to accomplish a certain purpose.
Hence,
in such applications of quantum mechanics,
an experimenter has to be able to
realize the quantum measurement described by a pre-designed POVM
with any desired accuracy.
Therefore the pre-designed POVM has to be \textit{computable}.
In the previous work \cite{T03},
we investigated what appears in the framework of quantum mechanics
if we take into account the computability of a POVM
for a finite dimensional quantum system.
We obtained a new kind of inequalities of quantum mechanics about
the probability of each measurement outcome in a computable POVM measurement
performed upon a finite dimensional quantum system.
In order to derive these inequalities,
we introduced the notion of a universal semi-POVM
on a finite dimensional quantum system,
as a generalization of the universal probability to a matrix-valued function.
The present work is, in essence,
an extension of the work \cite{T03} to infinite dimensional setting
with respect to the form of the theory.

The first attempt to extend the universal probability to an operator
is done by \cite{G01} for finite dimensional quantum system.
The purpose of \cite{G01} is mainly
to define the information content of an individual pure quantum state,
i.e.,
to define the \textit{quantum Kolmogorov complexity} of the quantum state,
while such an attempt is not the purpose of
both \cite{T03} and the present paper.
\cite{G01} generalized the universal probability to
a matrix-valued function $\boldsymbol{\mu}$,
called the \textit{quantum universal semi-density matrix}.
The function $\boldsymbol{\mu}$
maps any positive integer $N$ to
an $N\times N$ positive semi-definite Hermitian matrix $\boldsymbol{\mu}(N)$
with its trace less than or equal to one.
\cite{G01} proposed to regard $\boldsymbol{\mu}(N)$
as an analogue of a density matrix of a quantum system
whose state space has finite dimension $N$.
Since the dependency of $\boldsymbol{\mu}(N)$ on $N$ is
crucial to the framework of \cite{G01},
it would not seem clear how to extend the framework of \cite{G01}
to an infinite dimensional quantum system.
By comparison, the extension is clear to our framework.

In quantum mechanics,
what is represented by an operator is
either a quantum state or a measurement operator.
In \cite{T03} and the present work
we generalize the universal probability to an operator-valued function
in different way from \cite{G01}, and identify it with an analogue of a POVM.
We do not stick to defining the information content of a quantum state.
Instead, we focus our thoughts on
properly extending algorithmic information theory to quantum region
while keeping an appealing feature of the theory.

\subsection{Organization of the paper}

We begin in Section \ref{preliminaries}
with some basic notation and the results of algorithmic information theory.
In Section \ref{infinite dimension},
we introduce our definition of universal semi-POVM
after considering mathematical constraints on it.
We then propose our extension of $\Omega$
to an operator in infinite dimensional quantum system in Section \ref{omega}.
The introduction of universal semi-POVM
also enables us to extend $\K(s)$ to an operator
in a Hilbert space of infinite dimension.
In Section \ref{ovait},
we introduce the extension of $\K(s)$
and study its properties.
We conclude this paper
with a discussion about the future direction of our work
in Section \ref{discussion}.

\section{Preliminaries}
\label{preliminaries}

\subsection{Notation}

We start with some notation about numbers and matrices
which will be used in this paper.

$\#S$ is the cardinality of $S$ for any set $S$.
$\N \equiv \left\{0,1,2,3,\dotsc\right\}$ is the set of natural numbers,
and $\N^+$ is the set of positive integers.
$\Q$ is the set of rational numbers.
$\R$ is the set of real numbers, and $\C$ is the set of complex numbers.
$\CQ$ is the set of the complex numbers in the form of $a+ib$ with
$a,b\in\Q$.
For any matrix $A$, $A^\dagger$ is the adjoint of $A$.
Let $N\in\N^+$.
$\C^N$ is the set of column vectors consisting of $N$ complex numbers.
$\Her(N)$ is the set of $N\times N$ Hermitian matrices.
For each $A\in\Her(N)$, the \textit{norm} of $A$ is denoted by $\|A\|$,
i.e., $\|A\|=\max\{\abs{\nu}\mid\nu\text{ is an eigenvalue of }A\}$.
For each $A, B\in\Her(N)$,
we write $A\leqslant B$ if $B-A$ is positive semi-definite.
$\HerQ(N)$ is the set of $N\times N$ Hermitian matrices
whose elements are in $\CQ$.
$\diag(x_1,\dots,x_N)$ is the diagonal matrix whose $(j,j)$-element is $x_j$.

\subsection{Algorithmic information theory}
\label{ait}

In the following
we concisely review some definitions and results of
algorithmic information theory \cite{C75,C87a}.
We assume that the reader is familiar with algorithmic information theory
in addition to the theory of computable analysis.
(See e.g.~Chapter 0 of \cite{PR89} for the treatment of
the computability of complex numbers and complex functions on a discrete set.)

$\X \equiv
\left\{
  \lambda,0,1,00,01,10,11,000,001,010,\dotsc
\right\}$
is the set of finite binary strings
where $\lambda$ denotes the \textit{empty string},
and $\X$ is ordered as indicated.
We identify any string in $\X$ with a positive integer in this order,
i.e.,
we consider $\varphi\colon \X\to\N^+$ such that $\varphi(s)=1s$
where the concatenation $1s$ of strings $1$ and $s$ is regarded
as a dyadic integer,
and then we identify $s$ with $\varphi(s)$.
For any $s \in \X$, $\abs{s}$ is the \textit{length} of $s$.
A subset $S$ of $\X$ is called a \textit{prefix-free set}
if no string in $S$ is a prefix of another string in $S$.

A \textit{computer} is a partial recursive function
$C\colon \X\to \X$ whose domain of definition
is a prefix-free set.
For each computer $C$ and each $s \in \X$,
$\K_C(s)$ is defined by
$\K_C(s) \equiv
\min
\left\{\,
  \abs{p}\,\big|\;p \in \X\>\&\>C(p)=s
\,\right\}$.
A computer $U$ is said to be \textit{optimal} if
for each computer $C$ there exists a constant $\Sim(C)$
with the following property;
if $C(p)$ is defined, then there is a $p'$ for which
$U(p')=C(p)$ and $\abs{p'}\le\abs{p}+\Sim(C)$.
It is then shown that there exists
an optimal computer.
We choose any one optimal computer $U$ as the standard one for use,
and define $\K(s) \equiv \K_U(s)$,
which is referred to as
the \textit{program-size complexity} of $s$,
the \textit{information content} of $s$, or
the \textit{Kolmogorov complexity} of $s$
\cite{G74,L74,C75}.

Let $V$ be any optimal computer.
For any $s\in \X$, $P_V(s)$ is defined as $\sum_{V(p)=s}2^{-\abs{p}}$.
Chaitin's halting probability $\Omega_V$ of $V$ is defined by
\begin{equation}\label{Chaitin's omega}
  \Omega_V\equiv\sum_{V(p)\text{ is defined}}2^{-\abs{p}}.
\end{equation}
For any $\alpha\in(0,1]$,
we say that $\alpha$ is \textit{random} if
there exists $c\in\N$ such that, for any $n\in\N^+$, $n-c\le \K(\alpha_n)$
where $\alpha_n$ is the first $n$ bits of the base-two expansion of $\alpha$.
Then \cite{C75} showed that,
for any optimal computer $V$, $\Omega_V$ is random.
It is shown that $0<\Omega_V<1$ for any optimal computer $V$.

The class of computers is equal to the class of functions
which are computed by \textit{self-delimiting Turing machines}.
A self-delimiting Turing machine is a deterministic Turing machine
which has two tapes, a program tape and a work tape.
The program tape is infinite to the right,
while the work tape is infinite in both directions.
The program tape is read-only and
the tape head of the program tape cannot move to the left.
On the other hand,
the work tape is read/write and the tape head of the work tape
can move in both directions.
A self-delimiting Turing machine computes a partial function $f\colon\X\to\X$
as follows.
The machine starts in the initial state with
an input binary string $s$ on its program tape and the work tape blank.
The left-most cell of the program tape is blank
and the tape head of the program tape initially scans this cell.
The input string lies immediately to the right of this cell.
If the machine eventually halts with the tape head of the program tape
scanning the last bit of the input string $s$,
then $f(s)$ is defined as the string extending to the right
from the cell of the work tape which is being scanned to the first blank cell.
Otherwise, $f(s)$ is not defined.
Since the computation must end with the tape head of the program tape
scanning the last bit of the input string $s$
whenever $f(s)$ is defined,
the domain of definition of $f$ is a prefix-free set.
A self-delimiting Turing machine is called \textit{universal} if
it computes an optimal computer.
Let $M_V$ be a universal self-delimiting Turing machine which
computes an optimal computer $V$.
Then $P_V(s)$ is the probability that $M_V$ halts and outputs $s$
when $M_V$ starts on the program tape filled with an infinite binary string
generated by infinitely repeated tosses of a fair coin.
Therefore $\Omega_V=\sum_{s\in\X} P_V(s)$ is the probability that
$M_V$ just halts
under the same setting.
\cite{C75} showed the following theorem.

\begin{theorem}\label{eup}
  For any optimal computer $V$,
  both $2^{-\K_V(s)}$ and $P_V(s)$ are universal probabilities.
\end{theorem}

By Theorem \ref{eup}, we see that, for any universal probability $m$,

\begin{equation}\label{eq: K_m}
  \K(s)=-\log_2 m(s)+O(1).
\end{equation}

Thus it is possible to define $\K(s)$ as $-\log_2 m(s)$
with any one universal probability $m$ instead of as $\K_U(s)$.
Note that
the difference up to an additive constant is inessential to
algorithmic information theory.
Any universal probability is not computable,
as corresponds to the uncomputability of $\K(s)$.
As a result, we see that
$0<\sum_{s\in\X}m(s)<1$ for any universal probability $m$.

We can give another characterization of $\Omega_V$
using a universal probability,
as seen in the following theorem.
The proof of the theorem is based on
Theorem \ref{eup} above and the result of \cite{CHKW01}.
\begin{theorem}\label{omega-equiv-universal}
  For any $\alpha\in\R$,
  $\alpha=\sum_{s\in\X}m(s)$ for some universal probability $m$
  if and only if $\alpha=\Omega_V$
  for some optimal computer $V$.
\end{theorem}

\begin{proof}
  The ``if'' part follows
  from Theorem \ref{eup} and $\Omega_V=\sum_{s\in\X}P_V(s)$.
  The proof of the ``only if'' part is as follows.
  We say an increasing converging computable sequence $\{a_n\}$ of
  rational numbers is \textit{universal} if
  for every increasing converging computable sequence $\{b_n\}$ of
  rational numbers,
  there exists a real number $c>0$ such that, for all $n\in\N^+$,
  $c(\alpha-a_n)\ge \beta-b_n$
  where $\alpha=\lim_{n\to\infty}a_n$ and $\beta=\lim_{n\to\infty}b_n$.
  Theorem 6.6 in \cite{CHKW01} shows that,
  for any $\alpha\in(0,1)$,
  $\alpha=\Omega_V$ for some optimal computer $V$ if and only if
  there exists a universal increasing computable sequence of rational numbers
  which converges to $\alpha$.
  Thus it is sufficient to show that
  there exists a universal increasing computable sequence of rational numbers
  converging to $\sum_{s\in\X}m(s)$.
  Since $m$ is a lower-computable semi-measure,
  there exists a total recursive function
  $f\colon\N^+\times \X\to\Q$ such that, for each $s\in \X$,
  $\lim_{n\to\infty} f(n,s)=m(s)$ and
  $\forall\,n\in\N^+\;\>0\le f(n,s)\le f(n+1,s)$.
  We define an increasing computable sequence $\{a_n\}$ of rational numbers
  by $a_n=\sum_{s=1}^n f(n,s)$.
  Then we have
  $\abs{a_n-\sum_{s\in\X}m(s)}\le
  \sum_{s=1}^l\abs{f(n,s)-m(s)}+\sum_{s=l+1}^\infty m(s)$
  for any $l,n\in\N^+$ with $l<n$.
  Thus, by considering sufficiently large $n$
  for each sufficiently large $l$,
  we see that $\lim_{n\to\infty}a_n=\sum_{s\in\X}m(s)$.
  Let $\{b_n\}$ be an increasing computable sequence of rational numbers
  converging to $\beta$.
  We define $r\colon \X\to\Q\cap[0,\infty)$ by
  $r(s)=(b_{s}-b_{s-1})/d$ for any $s>1$ and $r(1)=0$,
  where $d$ is any one positive integer with $\beta-b_1\le d$.
  Then we see that $\sum_{s\in\X}r(s)=(\beta-b_1)/d\le 1$ and
  $r$ is a total recursive function.
  Therefore $r$ is a lower-computable semi-measure.
  Thus there exists a $c>0$ such that $cr(s)\le m(s)$ for all $s\in\X$.
  Hence we have
  $c(\beta-b_n)/d\le\sum_{s=n+1}^\infty m(s)=
  \sum_{s=1}^\infty m(s)-\sum_{s=1}^n m(s)$ and therefore
  $\beta-b_n\le d/c(\sum_{s\in\X}m(s)-a_n)$.
  Thus the proof is completed.
\end{proof}
In the present paper,
we extend a universal probability to a semi-POVM on $\X$.
Thus,
Theorem \ref{omega-equiv-universal} suggests that
an extension of $\Omega_V$ to an operator can be defined
as the sum of the POVM elements of such a semi-POVM on $\X$.
Therefore the most important thing is how to extend
a universal probability to a semi-POVM on $\X$
on a Hilbert space of infinite dimension.
We do this first in what follows.

\section{Extension of universal probability}
\label{infinite dimension}

In order to extend a universal probability to a semi-POVM on $\X$
which operates
on an infinite dimensional Hilbert space,
we have to develop a theory of computability for points and operators
of such a space.
We can construct the theory on any concrete Hilbert spaces
such as $l^2$ and $L^2(\R^{3n})$ with $n\in\N^+$
(the latter represents the state space of $n$ quantum mechanical particles
moving in three-dimensional space).
For the purpose of generality, however,
we here adopt an axiomatic approach which encompasses a variety of spaces.
Thus we consider
the notion of a \textit{computability structure on a Banach space}
which was introduced by \cite{PR89} in the late 1980s.

\subsection{Computability structures on a Banach space}

Let $X$ be a complex Banach space with a norm $\|\cdot\|$,
and let $\fmi$ be a nonempty set of sequences in $X$.
We say $\fmi$ is a \textit{computability structure} on $X$
if the following three axioms;
Axiom \ref{axiom1}, \ref{axiom2}, and \ref{axiom3} hold.
A sequence in $\fmi$ is regarded as a \textit{computable sequence} in $X$.

\begin{axiom}[Linear Forms]\label{axiom1}
  Let $\{x_{n}\}$ and $\{y_{n}\}$ be in $\fmi$,
  let $\{\alpha_{nk}\}$ and $\{\beta_{nk}\}$ be 
  computable double sequences of complex numbers,
  and let $d\colon\N^+\to\N^+$ be a total recursive function.
  Then the sequence
  \begin{equation*}
    s_n=\sum_{k=1}^{d(n)}(\alpha_{nk}x_k+\beta_{nk}y_k)
  \end{equation*}
  is in $\fmi$.
\end{axiom}

For any double sequence $\{x_{nm}\}$ in $X$,
we say $\{x_{nm}\}$ is \textit{computable} with respect to $\fmi$ if
it is mapped to a sequence in $\fmi$ by any one recursive bijection
from $\N^+$ to $\N^+\times\N^+$.
An element $x\in X$ is called \textit{computable} with respect to $\fmi$
if the sequence $\{x,x,x,\dotsc\}$ is in $\fmi$.

\begin{axiom}[Limits]\label{axiom2}
  Suppose that
  a double sequence $\{x_{nm}\}$ in $X$ is computable with respect to $\fmi$,
  $\{y_n\}$ is a sequence in $X$,
  and
  there exists a total recursive function
  $e\colon\N^+\times\N^+\to\N^+$
  such that $\|x_{ne(n,k)}-y_n\|\le 2^{-k}$ for all $n,k\in\N^+$.
  Then $\{y_{n}\}$ is in $\fmi$.
\end{axiom}

\begin{axiom}[Norms]\label{axiom3}
  If $\{x_{n}\}$ is in $\fmi$,
  then the norms $\{\|x_{n}\|\}$ form a computable sequence of real numbers.
\end{axiom}

We say a sequence $\{e_{n}\}$ in $X$ is a \textit{generating set} for $X$ or
a \textit{basis} for $X$ if
the set of all finite linear combinations of the $e_{n}$ is dense in $X$.

\begin{definition}
  Let $X$ be a Banach space with a computability structure $\fmi$.
  We say the pair $(X,\fmi)$ is effectively separable
  if there exists a sequence $\{e_{n}\}$ in $\fmi$ which
  is a generating set for $X$.
  Such a sequence $\{e_{n}\}$ is called
  an effective generating set for $(X,\fmi)$
  or a computable basis for $(X,\fmi)$.
\end{definition}

Throughout the rest of this paper,
we assume that
$X$ is an arbitrary complex Hilbert space
of infinite dimension with a computability structure $\fmi$
such that $(X,\fmi)$ is effectively separable.
We choose any one such a computability structure $\fmi$ on $X$
as the standard one throughout the rest of this paper,
and we do not refer to $\fmi$ hereafter.
For example,
we will simply say a sequence $\{x_{n}\}$ is computable
instead of saying $\{x_{n}\}$ is in $\fmi$.

We next define a notion of computability for a semi-POVM on $\X$
as a natural extension of
the notion of an \textit{effectively determined} bounded operator
which is defined in \cite{PR89}.
\begin{definition}[computability of semi-POVM]\label{computable-povm-infinite}
  Let $R$ be a semi-POVM on $\X$.
  We say $R$ is computable
  if there exists an effective generating set $\{e_{n}\}$ for $X$ such that
  the mapping $(s,n)\longmapsto (R(s))e_{n}$
  is a computable double sequence in $X$.
\end{definition}
Recall that we identify $\X$ with $\N^+$ in this paper.
For any semi-POVM $R$ on $\X$,
based on Axiom \ref{axiom1}, \ref{axiom2}, \ref{axiom3},
and $\|R(s)\|\le 1$ for all $s\in\X$,
we can show that
if $R$ is computable
then $\{(R(s))e_{n}\}$ is a computable double sequence in $X$
for every effective generating set $\{e_{n}\}$ for $X$.

The following two lemmas are frequently used
throughout the rest of this paper.

\begin{lemma}\label{bounded monotonic sequence}
  Let $\{A_n\}$ be a sequence of operators in $\HO$.
  Suppose that
  there exists a $B\in\HO$ such that,
  for all $n$, $A_n\leqslant A_{n+1}\leqslant B$.
  Then there exists an $A\in\HO$
  such that $\{A_n\}$ converges strongly to $A$ as $n\to\infty$
  and $A\leqslant B$.
\end{lemma}

The proof of Lemma \ref{bounded monotonic sequence} is
given at Section 104 of \cite{RS90}.

\begin{lemma}\label{monotone hasamiuti}
  Let $\{A_n\}$ and $\{B_n\}$ be sequences of operators in $\HO$.
  Suppose that
  (i) $A_n\leqslant B_n \leqslant A_{n+1}$ for all $n$, and
  (ii) $\{A_n\}$ converges strongly to some $A\in\HO$ as $n\to\infty$.
  Then $\{B_n\}$ also converges strongly to $A$ as $n\to\infty$.
\end{lemma}

\begin{proof}
  Since $A_n\leqslant A$ for all $n$,
  $B_n\leqslant B_{n+1}\leqslant A$ for all $n$.
  It follows from Lemma \ref{bounded monotonic sequence}
  that there exists a $B\in\HO$
  to which $\{B_n\}$ converges strongly as $n\to\infty$.
  Note that, for any $x\in X$,
  $\ip{A_nx}{x}\le \ip{B_nx}{x}\le \ip{A_{n+1}x}{x}$.
  Thus $\ip{Bx}{x}=\ip{Ax}{x}$ for any $x\in X$,
  and therefore we have $B=A$.
  This completes the proof.
\end{proof}

\subsection{Universal semi-POVM}

We first introduce
the notion of a \textit{lower-computable} semi-POVM on $\X$,
which is an extension of the notion of a lower-computable semi-measure
over a semi-POVM on $\X$.
Our definition of a lower-computable semi-POVM premises
the following lemma proved in \cite{PR89}.
We say a basis $\{e_{n}\}$ for $X$ is \textit{orthonormal} if
$\ip{e_{m}}{e_{n}}=\delta_{mn}$ for any $m,n\in\N^+$.

\begin{lemma}[Pour-El and Richards \cite{PR89}]\label{econs}
  Let $Y$ be a Hilbert space with a computability structure $\phi$
  such that $(Y,\phi)$ is effectively separable.
  Then there exists a computable orthonormal basis for $(Y,\phi)$.
\end{lemma}

By the above lemma,
we are given free access to the use of a computable orthonormal basis for $X$
in what follows.
The following definition is also needed to introduce
the notion of a lower-computable semi-POVM on $\X$.

\begin{definition}\label{def-approximation-from-below}
  Let $\{e_i\}$ be an orthonormal basis for $X$.
  For any $T\in\BO$ and $m\in\N^+$,
  we say $T$ is an $m$-square operator on $\{e_i\}$ if
  for all $k,l\in\N^+$ if $k>m$ or $l>m$ then $\ip{Te_k}{e_l}=0$.
  Furthermore, we say $T$ is an $m$-square rational operator on $\{e_i\}$
  if $T$ is an $m$-square operator on $\{e_i\}$ and for all $k,l\in\N^+$,
  $\ip{Te_k}{e_l}\in\CQ$
\end{definition}

The following Lemma \ref{positivity-condition} is suggestive to
fix the definition of a lower-computable semi-POVM on $\X$.
By Lemma \ref{positivity-condition},
we can effectively check whether $S \leqslant T$ holds or not,
given $S,T\in\HO$ and $m\in\N^+$ such that
$S$ and $T$ are $m$-square operators on an orthonormal basis for $X$.

\begin{lemma}\label{positivity-condition}
  Let $T\in\HO$,
  and let $\{e_i\}$ be an orthonormal basis for $X$.
  Then, the following three conditions (i), (ii), and (iii) are
  equivalent to one another.
  \begin{enumerate}
    \item $T$ is a positive operator.
    \item For every $m\in\N^+$,
      \begin{equation*}
        \left(
        \begin{array}{ccc}
          \ip{Te_{1}}{e_{1}} & \dotsb & \ip{Te_{1}}{e_{m}} \\
          \vdots & & \vdots \\
          \ip{Te_{m}}{e_{1}} & \dotsb & \ip{Te_{m}}{e_{m}}
        \end{array}
        \right) \geqslant 0.
      \end{equation*}
    \item For every finite sequence
      $\nu_1,\dots,\nu_m\in\N^+$ with $\nu_1<\dots<\nu_m$,
      \begin{equation*}
        \det
        \left(
        \begin{array}{ccc}
          \ip{Te_{\nu_1}}{e_{\nu_1}} & \dotsb & \ip{Te_{\nu_1}}{e_{\nu_m}} \\
          \vdots & & \vdots \\
          \ip{Te_{\nu_m}}{e_{\nu_1}} & \dotsb & \ip{Te_{\nu_m}}{e_{\nu_m}}
        \end{array}
        \right) \ge 0.
      \end{equation*}
  \end{enumerate}
\end{lemma}

\begin{proof}
  We note the elementary result of linear algebra that,
  for any $A\in\Her(N)$, $0\leqslant A$ if and only if
  all principal minors of $A$ are non-negative.
  Thus the conditions (ii) and (iii) are equivalent.
  We show the equivalence between the conditions (i) and (ii).
  For each $m\in\N^+$, let $V_m=\C e_1+\dots+\C e_m$.
  Then, for every $x\in V_m$, we see that $\ip{Tx}{x}\ge 0$ if and only if
  $\sum_{i,j=1}^m c_i\ip{Te_{i}}{e_{j}}\overline{c_j}\ge 0$
  where $\{c_i\}$ satisfies that $x=\sum_{i=1}^{m} c_i e_i$.
  Thus, the condition (ii) is equivalent to the condition that,
  for any $m\in\N^+$ and any $x\in V_m$, $\ip{Tx}{x}\ge 0$.
  Since $T\in\BO$,
  the latter condition is further equivalent to the condition (i).
  Hence, the proof is completed.
\end{proof}

We recall that, for any lower-computable semi-measure $r$,
there exists a total recursive function $f\colon\N^+\times \X\to\Q$
such that, for each $s\in \X$,
$\lim_{n\to\infty} f(n,s)=r(s)$ and
$\forall\,n\in\N^+\;\>0\le f(n,s)\le f(n+1,s)\le r(s)$.
We here consider how to extend this $f$ to an operator
in order to define a lower-computable semi-POVM $R$ on $\X$.
Let $\{e_i\}$ be an orthonormal basis for $X$.
When we prove the existence of a universal semi-POVM
(i.e., Theorem \ref{existence-of-universal-semi-POVM})
below,
especially in the proof of Lemma \ref{universal-generator},
we have to be able to decide whether $f(n,s)\leqslant f(n+1,s)$
in the sequence $\{f(n,s)\}_{n\in\N^+}$ of operators which converges to $R(s)$.
Thus, firstly, it is necessary for each $f(s,n)$
to be an $m$-square rational operator on $\{e_i\}$ for some $m\in\N^+$.
If so we can use Lemma \ref{positivity-condition}
to check $f(n,s)\leqslant f(n+1,s)$.
On that basis,
in order to complete the definition of a lower-computable semi-POVM,
it seems at first glance that we have only to require
that $0\leqslant f(n,s)\leqslant f(n+1,s)\leqslant R(s)$
and $f(n,s)$ converges to $R(s)$ in an appropriate sense.
Note that each operator $f(n,s)$ in the sequence has to be positive
in order to guarantee that the limit $R(s)$ is positive.
However,
this passing idea does not work properly
as shown by the following consideration.

For simplicity,
we consider matrices in $\Her(N)$ with $N\ge 2$ instead of operators in $X$.
We show that for some computable matrix $A\geqslant 0$
there does not exist a total recursive function $F\colon\N^+\to\HerQ(N)$
such that
\begin{equation}\label{fiction}
  \lim_{n\to\infty} F(n)=A \quad\;\text{and}\quad\;
  \forall\,n\in\N^+\;\;0\leqslant F(n)\leqslant A.
\end{equation}
This follows from Example \ref{2-square-matrix} below,
which is based on the following result of linear algebra.

\begin{proposition}\label{only-one-positive-eigenvalue}
  Let $A,B\in\Her(N)$.
  Suppose that $\rank A=1$ and $0\leqslant B\leqslant A$.
  Then $B=\tau A$ for some $\tau\in[0,1]$.
\end{proposition}

\begin{proof}
  Since $A\in\Her(N)$ and $\rank A=1$,
  there exist an $N\times N$ unitary matrix $U$ and a $\lambda>0$
  such that $A=U\diag(\lambda,0,\dots,0)\,U^{\dag}$.
  We write $U=(u_1\;u_2\;\dotsb\;u_N)$ with $u_k\in\C^N$.
  For each $k\ge 2$,
  since $u_k^{\dag}Au_k=0$ and $0\leqslant B\leqslant A$,
  we have $u_k^{\dag}Bu_k=0$.
  It follows from $0\leqslant B$ that $Bu_k=0$ for every $k\ge 2$.
  If $B$ has a nonzero eigenvalue $\nu$,
  then the eigenspace of $B$ corresponding to $\nu$ is $\C u_1$.
  Thus, we have $B=U\diag(\nu,0,\dots,0)\,U^{\dag}$ for some $\nu\in\R$.
  Since $0\le\nu=u_1^{\dag}Bu_1\le u_1^{\dag}Au_1=\lambda$,
  by setting $\tau=\lambda/\nu$, we have $B=\tau A$ and $\tau\in[0,1]$.
\end{proof}

\begin{example}\label{2-square-matrix}
  We consider the matrix $A\in\Her(2)$ given by
  \begin{equation*}
    A=
    \left(
    \begin{array}{cc}
      \frac{2}{3} & \frac{\sqrt{2}}{3} \\
      \frac{\sqrt{2}}{3} & \frac{1}{3}
    \end{array}
    \right).
  \end{equation*}
  First, we see that all elements of $A$ are computable real numbers,
  and therefore $A$ itself is computable.
  We can check that $\rank A=1$.
  In fact, $A$ has two eigenvalues $0$ and $1$.
  It can be shown that
  there does not exist any nonzero $B\in\HerQ(2)$ such that
  $0\leqslant B\leqslant A$.
  Contrarily, assume that such a $B$ exists.
  Then, by Proposition \ref{only-one-positive-eigenvalue},
  we have $B=\tau A$ for some $\tau\in(0,1]$, i.e.,
  \begin{equation*}
    B=
    \left(
    \begin{array}{cc}
      \frac{2}{3}\tau & \frac{\sqrt{2}}{3}\tau \\
      \frac{\sqrt{2}}{3}\tau & \frac{1}{3}\tau
    \end{array}
    \right).
  \end{equation*}
  However, for any $\tau>0$,
  it is impossible for all elements of $B$ to be simultaneously in $\CQ$. \qed
\end{example}

Thus, even in a non-effective manner,
we cannot get a sequence $\{F(n)\}\subset\HerQ(N)$ which satisfies
the condition \eqref{fiction}.
On the other hand,
for any positive semi-definite $A\in\Her(N)$ and any $n\in\N^+$,
there exists a $B\in\HerQ(N)$ such that
$0\leqslant B\leqslant A+2^{-n}E$,
where $E$ is the identity matrix.
This is because,
since $\HerQ(N)$ is dense in $\Her(N)$
with respect to the norm $\|\cdot\|$,
there exists a $B\in\HerQ(N)$ such that $\|A+2^{-n+1}/3E-B\|\le 2^{-n}/3$.
Thus we have $0\leqslant A+2^{-n}/3E\leqslant B\leqslant A+2^{-n}E$.
Furthermore we can show that, for any positive semi-definite $A\in\Her(N)$,
if $A$ is computable, then
there exists a total recursive function $F\colon\N^+\to\HerQ(N)$ such that
(i) $\lim_{n\to\infty}F(n)=A$,
(ii) $0\leqslant F(n)$, and
(iii) $F(n)-2^{-n}E\leqslant F(n+1)-2^{-(n+1)}E\leqslant A$.
Note that a positive semi-definite matrix $A$ with rank $1$ 
as considered in Example \ref{2-square-matrix}
is not an atypical example as a POVM element
in quantum measurements,
since such a POVM element is common in a familiar projective measurement.

The foregoing consideration suggests the following definition of
a lower-computable semi-POVM on an infinite dimensional Hilbert space.
  
\begin{definition}\label{def-pre-lower-computable-semi-povm}
  Let $\{e_i\}$ be a computable orthonormal basis for $X$,
  and let $R$ be a semi-POVM on $\X$.
  We say $R$ is lower-computable with respect to $\{e_i\}$
  if there exist an $f\colon\N^+\times\X\to\PO$ and
  a total recursive function $g\colon\N^+\times\X\to\N^+$ such that
  \begin{enumerate}
    \item for each $s\in\X$,
      $f(n,s)$ converges strongly to $R(s)$ as $n\to\infty$,
    \item for all $n$ and $s$,
      $f(n,s)-2^{-n}I\leqslant f(n+1,s)-2^{-(n+1)}I$,
    \item for all $n$ and $s$, 
      $f(n,s)$ is a $g(n,s)$-square rational operator on $\{e_i\}$, and
    \item the mapping
      $\N^+\times\X\times\N^+\times\N^+\ni (n,s,i,j)
      \longmapsto \ip{f(n,s)e_i}{e_j}$
      is a total recursive function.
  \end{enumerate}
\end{definition}

In the above definition,
we choose the sequence $\{2^{-n}\}$ as the coefficients of $I$
in the inequality of the condition (ii).
However, by the following proposition,
we can equivalently replace $\{2^{-n}\}$ by
a general nonincreasing computable sequence of non-negative rational numbers
which converges to $0$.

\begin{proposition}\label{lower-computable-semi-povm-relaxed}
  Let $\{e_i\}$ be a computable orthonormal basis for $X$,
  and let $R$ be a semi-POVM on $\X$.
  Then,
  $R$ is lower-computable with respect to $\{e_i\}$
  if and only if there exist an $f'\colon\N^+\times\X\to\PO$,
  a total recursive function $g'\colon\N^+\times\X\to\N^+$, and
  a total recursive function $h\colon\N^+\times\X\to\Q$ such that
  \begin{enumerate}
    \item for each $s\in\X$,
      $f'(n,s)$ converges strongly to $R(s)$ as $n\to\infty$,
    \item for all $n$ and $s$,
      $f'(n,s)-h(n,s)I\leqslant f'(n+1,s)-h(n+1,s)I$,
    \item for each $s$, $\lim_{n\to\infty} h(n,s)=0$
      and $\forall\,n\in\N^+\;h(n,s)\ge h(n+1,s)\ge 0$,
    \item for all $n$ and $s$, 
      $f'(n,s)$ is a $g'(n,s)$-square rational operator on $\{e_i\}$, and
    \item the mapping
      $\N^+\times\X\times\N^+\times\N^+\ni (n,s,i,j)
      \longmapsto \ip{f'(n,s)e_i}{e_j}$
      is a total recursive function.
  \end{enumerate}
\end{proposition}

\begin{proof}
  The ``only if'' part is obvious, and we show the ``if'' part.
  To begin with,
  we define $\overline{h}(n,s)$ as $h(n,s)+2^{-n}$.
  It follows that
  $f'(n,s)-\overline{h}(n,s)I\leqslant f'(n+1,s)-\overline{h}(n+1,s)I$,
  $\lim_{n\to\infty} \overline{h}(n,s)=0$,
  and $\overline{h}(n,s)>\overline{h}(n+1,s)>0$.
  Without loss of generality,
  we assume that $\overline{h}(1,s)>1/2$.
  In what follows, we use the fact that,
  for any $A,B\in\HO$ and any $\alpha,\beta\in[0,1]$,
  if $A\leqslant B$ and $\alpha\le\beta$, then
  $A\leqslant (1-\alpha)A+\alpha B\leqslant (1-\beta)A+\beta B\leqslant B$.
  In order to define $f\colon\N^+\times\X\to\PO$
  and $g\colon\N^+\times\X\to\N^+$
  which satisfy the conditions (i), (ii), (iii), and (iv)
  in Definition \ref{def-pre-lower-computable-semi-povm},
  we follow the procedure below for each $s$.
  Initially we set $m:=1$ and $n:=1$.
  
  Assume that $f(k,s)$ and $g(k,s)$ have so far been defined
  for all $k\in\{1,\dots,n-1\}$.
  We look for the least $l>m$ with $2^{-n}\ge \overline{h}(l,s)$.
  Since $\lim_{k\to\infty} \overline{h}(k,s)=0$, we can find such an $l$.
  Once we get the $l$,
  we calculate the finite set
  $S=
  \{
    k\in\N^+\mid k\ge n\;\&\;\overline{h}(m,s)>2^{-k}\ge \overline{h}(l,s)
  \}$.
  For each $k\in S$,
  we then define $f(k,s)$ as $(1-\alpha_k)f'(m,s)+\alpha_k f'(l,s)$ where
  $\alpha_k=(\overline{h}(m,s)-2^{-k})/(\overline{h}(m,s)-\overline{h}(l,s))$,
  and we also define $g(k,s)$ as $\max\{g'(m,s),g'(l,s)\}$.
  It follows that, for every $k\in S-\{n\}$,
  \begin{equation*}
    f'(m,s)-\overline{h}(m,s)I\leqslant
    f(k-1,s)-2^{-(k-1)}I\leqslant
    f(k,s)-2^{-k}I\leqslant
    f'(l,s)-\overline{h}(l,s)I
  \end{equation*}
  and $f(k,s)$ is a $g(k,s)$-square rational operator on $\{e_i\}$.
  We then set $m:=l$ and $n:=n+\#S$, and repeat this procedure.
  
  It can be checked that the $f$ and $g$ defined by this procedure
  satisfy the desired properties.
  Especially,
  in a similar manner to the proof of Lemma \ref{monotone hasamiuti}
  we can show that, for each $s\in\X$,
  $f(n,s)$ converges strongly to $R(s)$ as $n\to\infty$.
  Thus the proof is completed.
\end{proof}

In Proposition \ref{lower-computable-semi-povm-independent} below,
we show that the lower-computability of a semi-POVM on $\X$
given in Definition \ref{def-pre-lower-computable-semi-povm}
does not depend on the choice of a computable orthonormal basis
used in the definition.
The proof of Proposition \ref{lower-computable-semi-povm-independent}
uses the following Lemma \ref{i-to-im},
which follows from the equivalence between the conditions (i) and (iii)
in Lemma \ref{positivity-condition}.

\begin{lemma}\label{i-to-im}
  Let $T\in\HO$ be an $m$-square operator on 
  an orthonormal basis $\{e_i\}$ for $X$.
  For any real number $a>0$,
  $0\leqslant T+aI$ if and only if $0\leqslant T+aI_m$
  where $I_m$ is the operator in $\HO$ such that
  $I_m e_i=e_i$ if $i\le m$ and $I_m e_i=0$ otherwise.
\end{lemma}

By Lemma \ref{i-to-im},
in order to check whether the condition (ii) of
Definition \ref{def-pre-lower-computable-semi-povm} holds,
we can equivalently check the condition that
$0\leqslant f(n+1,s)-f(n,s)+2^{-n-1}I_m$
if $f(n,s)$ and $f(n+1,s)$ are
$m$-square operators on an orthonormal basis $\{e_i\}$ for $X$.

For each $T\in\BO$, the \textit{norm} of $T$ is denoted by $\|T\|$. 
Throughout the rest of this paper,
we will frequently use the property:
For any $\varepsilon\ge 0$ and any $T\in\HO$,
$\|T\|\le\varepsilon$ if and only if
$-\varepsilon I \leqslant T \leqslant \varepsilon I$.
For each $T\in\BO$, we define $\|T\|_2$ as
$(\sum_{i=1}^\infty \|Te_i\|^2)^{1/2}\in[0,\infty]$,
where $\{e_{n}\}$ is an arbitrary orthonormal basis for $X$.
Note that $\|T\|_2$ is independent of
the choice of an orthonormal basis $\{e_{n}\}$ for $X$,
and $\|T\|\le\|T\|_2$.
These properties of $\|\cdot\|_2$ are used
in the proof of Proposition \ref{lower-computable-semi-povm-independent}.

\begin{proposition}\label{lower-computable-semi-povm-independent}
  Let $R$ be a semi-POVM on $\X$,
  and let $\{e_i\}$ and $\{e'_k\}$ be computable orthonormal bases for $X$.
  Then,
  $R$ is lower-computable with respect to $\{e_i\}$ if and only if
  $R$ is lower-computable with respect to $\{e'_k\}$.
\end{proposition}

\begin{proof}
  We first define $u_{ki}=\ip{e'_k}{e_i}$.
  Then $\{u_{ki}\}$ is the computable double sequence of complex numbers
  which satisfies
  $e'_k=\sum_{i=1}^\infty u_{ki}e_i$.
  Assume that $R$ is lower-computable with respect to $\{e_i\}$.
  Then there exist an $f\colon\N^+\times\X\to\PO$ and
  a total recursive function $g\colon\N^+\times\X\to\N^+$ which satisfy
  the conditions (i), (ii), (iii), and (iv)
  in Definition \ref{def-pre-lower-computable-semi-povm}.
  In what follows,
  we show that $R$ is lower-computable with respect to $\{e'_i\}$.
  To begin with,
  we note that
  $\sum_{i,j=1}^{g(n,s)} \abs{\ip{f(n,s)e_i}{e_j}}^2={\|f(n,s)\|_2}^2=
  \sum_{k,l=1}^{\infty} \abs{\ip{f(n,s)e'_k}{e'_l}}^2$.
  Here, since
  $\ip{f(n,s)e'_k}{e'_l}=\sum_{i,j=1}^{g(n,s)}u_{ki}\ip{f(n,s)e_i}{e_j}
  \overline{u_{lj}}$,
  $\{\ip{f(n,s)e'_k}{e'_l}\}$ is
  a computable fourfold sequence of complex numbers.
  Thus, there exists a total recursive function
  $g'\colon\N^+\times\X\to\N^+$ such that
  \begin{equation*}
    \abs{\sum_{i,j=1}^{g(n,s)} \abs{\ip{f(n,s)e_i}{e_j}}^2-
    \sum_{k,l=1}^{g'(n,s)} \abs{\ip{f(n,s)e'_k}{e'_l}}^2} \le 2^{-2n-7}.
  \end{equation*}
  On the other hand,
  it is easy to show that
  there exists $\overline{f}\colon\N^+\times\X\to\HO$
  such that
  (i) for every $k,l\in\{1,\dots,g'(n,s)\}$,
  $\abs{\ip{(\overline{f}(n,s)-f(n,s))e'_k}{e'_l}}^2
  \le 1/g'(n,s)^2\,2^{-2n-7}$,
  (ii) $\overline{f}(n,s)$ is
  a $g'(n,s)$-square rational operator on $\{e'_k\}$, and
  (iii) the mapping $(n,s,k,l)\longmapsto\ip{\overline{f}(n,s)e'_k}{e'_l}$ is
  a total recursive function.
  Therefore we have
  \begin{eqnarray*}
    &&{\|\overline{f}(n,s)-f(n,s)\|_2}^2= \\
    &&\sum_{k,l=1}^{g'(n,s)}
    \abs{\ip{(\overline{f}(n,s)-f(n,s))e'_k}{e'_l}}^2+
    {\|f(n,s)\|_2}^2-
    \sum_{k,l=1}^{g'(n,s)} \abs{\ip{f(n,s)e'_k}{e'_l}}^2\\
    &&\le 2^{-2n-7}+2^{-2n-7}\le 2^{-2n-6}.
  \end{eqnarray*}
  Hence, $\|\overline{f}(n,s)-f(n,s)\|\le
  \|\overline{f}(n,s)-f(n,s)\|_2\le 2^{-n-3}$,
  and therefore $0\leqslant f(n,s)\leqslant \overline{f}(n,s)+2^{-n-3}I$.
  We then define $f'\colon\N^+\times\X\to\HO$
  by $f'(n,s)=\overline{f}(n,s)+2^{-n-3}I(n,s)$,
  where $I(n,s)\in\HO$ satisfies that
  $I(n,s)e'_k=e'_k$ if $k\le g'(n,s)$ and $I(n,s)e'_k=0$ otherwise.
  It follows that
  $f'(n,s)$ is a $g'(n,s)$-square rational operator on $\{e'_k\}$ and
  the mapping $(n,s,k,l)\longmapsto\ip{f'(n,s)e'_k}{e'_l}$
  is a total recursive function.
  In particular, by Lemma \ref{i-to-im},
  we have $0\leqslant f'(n,s)$.
  Since $\|f'(n,s)-f(n,s)\|\le
  \|\overline{f}(n,s)-f(n,s)\|+2^{-n-3}\|I(n,s)\|\le 2^{-n-2}$,
  $f'(n,s)-2^{-n-2}I\leqslant f(n,s)\leqslant f'(n,s)+2^{-n-2}I$.
  Using $f(n,s)-2^{-n}I\leqslant f(n+1,s)-2^{-(n+1)}I$,
  we have
  \begin{equation*}
    f(n,s)-2^{-(n-1)}I\leqslant f'(n,s)-(2^{-n-2}+2^{-n-1}+2^{-n})I
    \leqslant f(n+1,s)-2^{-n}I.
  \end{equation*}
  From this inequality,
  it is shown that
  \begin{equation*}
    f'(n,s)-(2^{-n-2}+2^{-n-1}+2^{-n})I
    \leqslant f'(n+1,s)-(2^{-n-3}+2^{-n-2}+2^{-n-1})I
  \end{equation*}
  and, for each $s\in\X$,
  $f'(n,s)$ converges strongly to $R(s)$ as $n\to\infty$.
  The latter follows from Lemma \ref{monotone hasamiuti}.
  Thus, by Proposition \ref{lower-computable-semi-povm-relaxed},
  $R$ is lower-computable with respect to $\{e'_k\}$.
  This completes the proof.
\end{proof}

Based on the above proposition,
we define the notion of a lower-computable semi-POVM on $\X$
independently of a choice of a computable orthonormal basis for $X$.

\begin{definition}[lower-computable semi-POVM on $\X$]
\label{def-lower-computable-semi-povm}
  Let $R$ be a semi-POVM on $\X$.
  We say $R$ is lower-computable if
  there exists a computable orthonormal basis $\{e_i\}$ for $X$
  such that $R$ is lower-computable with respect to $\{e_i\}$.
\end{definition}

Thus, for any semi-POVM $R$ on $\X$,
based on Proposition \ref{lower-computable-semi-povm-independent},
we see that
if $R$ is lower-computable then $R$ is lower-computable
with respect to every computable orthonormal basis for $X$.

Any computable function $r\colon\X\to[0,1]$ with $\sum_{s\in\X}r(s)\le 1$ is
shown to be a lower-computable semi-measure.
Corresponding to this fact
we can show Theorem \ref{computable-lower-computable} below.
In the theorem, however,
together with the computability of a semi-POVM $R$ on $\X$,
we need an additional assumption that
(i) each POVM element $R(s)$ is Hilbert-Schmidt and (ii) given $s$,
$\|R(s)\|_2$ can be computed to any desired degree of precision.
Here, for any $T\in\BO$,
we say $T$ is \textit{Hilbert-Schmidt} if $\|T\|_2<\infty$.
As an example,
consider a POVM $P$ on $\X$ with $(P(s))e_i=\delta_{si}e_i$,
where $\{e_i\}$ is a computable orthonormal basis for $X$.
Then $P$ is shown to be
a computable POVM on $\X$ which satisfies this additional assumption
(see the proof of Proposition \ref{difference}).
Note that the quantum measurement described by the $P$ is
a familiar projective measurement,
such as the measurement of the number of photons
in a specific mode of electromagnetic field.

\begin{theorem}\label{computable-lower-computable}
  Suppose that
  (i) $R\colon\X\to\BO$ is a computable semi-POVM on $\X$,
  (ii) $R(s)$ is Hilbert-Schmidt for every $s\in\X$,
  and (iii) $\{\|R(s)\|_2\}_{s\in\X}$ is a computable sequence of real numbers.
  Then $R$ is lower-computable.
\end{theorem}

\begin{proof}
  Let $\{e_i\}$ be any one computable orthonormal basis for $X$.
  Since
  $\{\ip{R(s)e_i}{e_j}\}$ is
  a computable triple sequence of complex numbers and
  $\{\|R(s)\|_2\}$ is a computable sequence of real numbers,
  it is easy to show that
  there exists a total recursive function
  $g\colon\N^+\times\X\to\N^+$ such that
  \begin{equation*}
    \abs{{\|R(s)\|_2}^2-
    \sum_{i,j=1}^{g(n,s)} \abs{\ip{R(s)e_i}{e_j}}^2}
    \le 2^{-2n-5}
  \end{equation*}
  and $g(n,s)\le g(n+1,s)$.
  Again,
  since
  $\{\ip{R(s)e_i}{e_j}\}$ is
  a computable triple sequence of complex numbers,
  we can show that
  there exists $\overline{f}\colon\N^+\times\X\to\HO$
  such that
  (i) for every $i,j\in\{1,\dots,g(n,s)\}$,
  $\abs{\ip{(R(s)-\overline{f}(n,s))e_i}{e_j}}^2\le 1/g(n,s)^2\,2^{-2n-5}$,
  (ii) $\overline{f}(n,s)$ is
  a $g(n,s)$-square rational operator on $\{e_i\}$, and
  (iii) the mapping $(n,s,i,j)\longmapsto\ip{\overline{f}(n,s)e_i}{e_j}$ is
  a total recursive function.
  Therefore we have
  \begin{eqnarray*}
    &&{\|R(s)-\overline{f}(n,s)\|_2}^2= \\
    &&\sum_{i,j=1}^{g(n,s)}
    \abs{\ip{(R(s)-\overline{f}(n,s))e_i}{e_j}}^2+
    {\|R(s)\|_2}^2-\sum_{i,j=1}^{g(n,s)} \abs{\ip{R(s)e_i}{e_j}}^2 \\
    &&\le 2^{-2n-5}+2^{-2n-5}\le 2^{-2n-4}.
  \end{eqnarray*}
  Hence,
  $\|R(s)-\overline{f}(n,s)\|\le\|R(s)-\overline{f}(n,s)\|_2\le 2^{-n-2}$,
  and therefore
  \begin{equation}\label{ineq-computable}
    \overline{f}(n,s)-2^{-n-2}I\leqslant R(s)
    \leqslant\overline{f}(n,s)+2^{-n-2}I.
  \end{equation}
  We then define $f\colon\N^+\times\X\to\HO$
  by $f(n,s)=\overline{f}(n,s)+2^{-n-2}I(n,s)$,
  where $I(n,s)\in\HO$ satisfies that
  $I(n,s)e_i=e_i$ if $i\le g(n,s)$ and $I(n,s)e_i=0$ otherwise.
  It follows that
  $f(n,s)$ is a $g(n,s)$-square rational operator on $\{e_i\}$ and
  the mapping $(n,s,i,j)\longmapsto\ip{f(n,s)e_i}{e_j}$
  is a total recursive function.
  In particular,
  by $0\leqslant R(s)$, the inequality \eqref{ineq-computable},
  and Lemma \ref{i-to-im},
  we have $0\leqslant f(n,s)$.
  It follows also from the inequality \eqref{ineq-computable} that
  $\|R(s)-f(n,s)\|\le\|R(s)-\overline{f}(n,s)\|+2^{-n-2}\|I(n,s)\|
  \le 2^{-n-1}$.
  Thus, for each $s\in\X$,
  $f(n,s)$ converges strongly to $R(s)$ as $n\to\infty$.
  Finally, we show that
  $f(n,s)-2^{-n}I\leqslant f(n+1,s)-2^{-(n+1)}I$.
  For that purpose,
  we note that
  $\overline{f}(n+1,s)-\overline{f}(n,s)\geqslant -(2^{-n-3}+2^{-n-2})I$ and
  $I(n,s)\leqslant I(n+1,s)\leqslant I$. The former follows from
  the inequality \eqref{ineq-computable}.
  Based on these inequalities, we have
  \begin{equation*}
    (f(n+1,s)-2^{-(n+1)}I)-(f(n,s)-2^{-n}I)\geqslant 2^{-n-3}(I-I(n+1,s))
    \geqslant 0.
  \end{equation*}
  This completes the proof.
\end{proof}

\begin{remark}
  It is open whether $R$ can be proved to be lower-computable
  only under the assumption that
  $R\colon\X\to\BO$ is a computable semi-POVM on $\X$.
\end{remark}

As a natural generalization of the notion of a universal probability,
the notion of a universal semi-POVM is defined as follows.

\begin{definition}[universal semi-POVM]\label{def-universal-semi-POVM}
  Let $M$ be a lower-computable semi-POVM on $\X$.
  We say that $M$ is a universal semi-POVM if
  for each lower-computable semi-POVM $R$ on $\X$,
  there exists a real number $c>0$ such that,
  for all $s\in \X$, $c\,R(s)\leqslant M(s)$.
\end{definition}

Most importantly we can show the existence of a universal semi-POVM.

\begin{theorem}\label{existence-of-universal-semi-POVM}
  There exists a universal semi-POVM.
\end{theorem}

In order to prove Theorem \ref{existence-of-universal-semi-POVM},
we need the following two lemmas.

\begin{lemma}\label{to-positive-semi-definite}
  Let $\{e_i\}$ be a computable orthonormal basis for $X$,
  and let $R$ be a semi-POVM on $\X$.
  If $R$ is lower-computable,
  then there exist an $f'\colon\N^+\times\X\to\PO$ and
  a total recursive function $g'\colon\N^+\times\X\to\N^+$ such that
  \begin{enumerate}
    \item the mapping
      $\displaystyle\X\ni s\longmapsto \frac{1}{2}R(s)+\frac{1}{2^{s+1}}I$ is
      a lower-computable semi-POVM on $\X$,
    \item for each $s\in\X$,
      $f'(n,s)$ converges strongly to
      $\displaystyle\frac{1}{2}R(s)+\frac{1}{2^{s+1}}I$ as $n\to\infty$,
    \item for all $n$ and $s$,
      $f'(n,s)\leqslant f'(n+1,s)$,
    \item for all $n$ and $s$, 
      $f'(n,s)$ is a $g'(n,s)$-square rational operator on $\{e_i\}$, and
    \item the mapping
      $\N^+\times\X\times\N^+\times\N^+\ni (n,s,i,j)
      \longmapsto\ip{f'(n,s)e_i}{e_j}$
      is a total recursive function.
  \end{enumerate}
\end{lemma}

\begin{proof}
  Since $R$ is lower-computable,
  there exist an $f\colon\N^+\times\X\to\PO$ and
  a total recursive function $g\colon\N^+\times\X\to\N^+$
  which satisfy the conditions (i), (ii), (iii), and (iv)
  in Definition \ref{def-pre-lower-computable-semi-povm}.
  Without loss of generality,
  we assume that $g(n,s)<g(n+1,s)$.
  For each $(n,s)\in\N\times\X$,
  let $I(n,s)$ be the operator in $\HO$ such that
  $I(n,s)e_i=e_i$ if $i\le g(n,s)$ and $I(n,s)e_i=0$ otherwise.
  Then we have $I(n,s)\leqslant I(n+1,s)$.
  It follows from
  $f(n,s)\leqslant f(n+1,s)+2^{-n-1}I$ and Lemma \ref{i-to-im}
  that $f(n,s)\leqslant f(n+1,s)+2^{-n-1}I(n+1,s)$.
  We define an $f'\colon\N^+\times\X\to\BO$
  by $f'(n,s)=1/2f(n+s,s)+2^{-s-1}(1-2^{-n})I(n+s,s)$,
  and define a total recursive function $g'\colon\N^+\times\X\to\N^+$
  by $g'(n,s)=g(n+s,s)$.
  Then we see that $0\leqslant f'(n,s)\leqslant f'(n+1,s)$.
  It is easy to check that
  $f'(n,s)$ is a $g'(n,s)$-square rational operator on $\{e_i\}$ and
  the mapping
  $\N^+\times\X\times\N^+\times\N^+\ni(n,s,i,j)\longmapsto\ip{f'(n,s)e_i}{e_j}$
  is a total recursive function.
  Since $I(n,s)$ converges strongly to $I$ as $n\to\infty$,
  $f'(n,s)$ converges strongly to $1/2R(s)+1/2^{s+1}I$.
  We have $\sum_{s\in\X} \{1/2R(s)+1/2^{s+1}I\}\leqslant
  1/2\sum_{s\in\X}R(s)+1/2I\leqslant I$.
  Thus,
  the mapping $\X\ni s\longmapsto 1/2R(s)+1/2^{s+1}I$ is
  a lower-computable semi-POVM on $\X$.
  This completes the proof.
\end{proof}

\smallskip

\begin{lemma}\label{universal-generator}
  Let $\{e_i\}$ be a computable orthonormal basis for $X$.
  Then there exist an $f\colon\N^+\times\N^+\times\X\to\PO$ and
  a total recursive function $g\colon\N^+\times\N^+\times\X\to\N^+$ such that
  \begin{enumerate}
    \item for all $l$, $n$, and $s$,
      $f(l,n,s)\leqslant f(l,n+1,s)$,
    \item for all $l$, $n$, and $s$, 
      $f(l,n,s)$ is a $g(l,n,s)$-square rational operator on $\{e_i\}$,
    \item the mapping
      $\N^+\times\N^+\times\X\times\N^+\times\N^+\ni (l,n,s,i,j)
      \longmapsto\ip{f(l,n,s)e_i}{e_j}$
      is a total recursive function,
    \item for each $l\in\N^+$,
      there exists a lower-computable semi-POVM $R_l$ on $\X$ such that,
      for every $s\in\X$,
      $f(l,n,s)$ converges strongly to $R_l(s)$ as $n\to\infty$, and
    \item for each lower-computable semi-POVM $R$ on $\X$,
      there exists an $l\in\N^+$ such that,
      for every $s\in\X$,
      $f(l,n,s)$ converges strongly to
      $\displaystyle\frac{1}{2}R(s)+\frac{1}{2^{s+1}}I$ as $n\to\infty$.
  \end{enumerate}
\end{lemma}

\begin{proof}
  We first note that, for any $A\in\HerQ(N)$,
  there exists a unique $T_A\in\HO$ such that
  $\ip{T_Ae_i}{e_j}=A_{ij}$ for every $i,j\in\N^+$ and
  $T_A$ is an $N$-square rational operator on $\{e_i\}$.
  
  Given $l\in\N^+$,
  for all $(n,s)\in\N^+\times\X$,
  $f(l,n,s)$ and $g(l,n,s)$ are defined through the following procedure.

  We first build the $l$-th Turing machine $M_l$.
  We make use of $M_l$
  as a machine which outputs
  a Hermitian matrix in $\bigcup_{N=1}^\infty \HerQ(N)$
  on an input $(n,s)\in\N^+\times\X$.
  Let $f_l\colon\N^+\times\X\to\bigcup_{N=1}^\infty \HerQ(N)$ be
  a partial recursive function computed by $M_l$ in this sense.
  For each $n\in\N^+$,
  let $S_n=\{(n-s+1,s)\mid s\in\X\;\&\;1\le s\le n\}$.
  In increasing order on $n$,
  we simulate the computations of $M_l$ on all inputs in $S_n$.
  During the procedure,
  we keep the function $h\colon\X\to\bigcup_{N=1}^\infty \HerQ(N)$
  and update it accordingly.
  For each $(n,s)\in\N^+\times\X$,
  $f(l,n,s)$ and $g(l,n,s)$ are defined
  as $T_{h(s)}$ and the order of the square matrix $h(s)$, respectively.
  Here $h(s)$ is one at the time step $n$ in the simulations.
  Initially we set $h(s):=0$ for all $s\in\X$ and $n:=1$.
  
  Assume that the simulations of $M_l$ on all inputs
  in $\bigcup_{k=1}^{n-1}S_{k}$ have so far been completed.
  We simulate the computations of $M_l$ on all inputs in $S_{n}$.
  If all such computations halt then we check whether the following
  three conditions hold:
  \begin{enumerate}
    \item $f_l(k,s)$ is defined for all $(k,s)\in S_n$,
    \item $T_{h(s)}\leqslant T_{f_l(k,s)}$ for all $(k,s)\in S_n$, and
    \item $\sum_{s=1}^nT_{f_l(n-s+1,s)}\leqslant I$.
  \end{enumerate}
  Note that we can effectively check
  whether the above conditions (ii) and (iii) hold,
  based on the equivalence between the conditions (i) and (iii)
  in Lemma \ref{positivity-condition}.
  If these three conditions hold then we set
  $h(s):=f_l(n-s+1,s)$ for each $s\in\{1,\dots,n\}$ and $n:=n+1$.
  We then repeat this procedure.
  
  We can show that the $f$ and $g$ defined by this procedure
  satisfy that
  (i) $0\leqslant f(l,n,s)\leqslant f(l,n+1,s)$,
  (ii) $f(l,n,s)$ is a $g(l,n,s)$-square rational operator on $\{e_i\}$, and
  (iii) the mapping
  $\N^+\times\N^+\times\X\times\N^+\times\N^+\ni (l,n,s,i,j)
  \longmapsto\ip{f(l,n,s)e_i}{e_j}$ and $g$ are total recursive functions.
  We also see that $\sum_{s=1}^mf(l,n,s)\leqslant I$ for any $l,m,n\in\N^+$.
  Thus we have $f(l,n,s)\leqslant I$ and therefore,
  by Lemma \ref{bounded monotonic sequence},
  there exists an $R_l\colon\X\to\PO$ such that
  $f(l,n,s)$ converges strongly to $R_l(s)$ as $n\to\infty$.
  Hence we have $\sum_{s=1}^mR_l(s)\leqslant I$.
  It follows from $0\leqslant R_l(s)$ and
  Lemma \ref{bounded monotonic sequence}
  that $ \sum_{s=1}^mR_l(s)$ converges strongly to
  $\sum_{s\in\X} R_l(s)\in\HO$ as $m\to\infty$ and
  $\sum_{s\in\X} R_l(s)\leqslant I$.
  Thus $R_l$ is a lower-computable semi-POVM on $\X$ for all $l$.
  
  Now, let $R$ be any lower-computable semi-POVM on $\X$.
  Then, by Lemma \ref{to-positive-semi-definite},
  there exist an $f'\colon\N^+\times\X\to\PO$ and
  a total recursive function $g'\colon\N^+\times\X\to\N^+$
  which satisfy the conditions (i), (ii), (iii), (iv), and (v) in the lemma.
  Based on the above construction of $f$,
  we see that there exists $k\in\N^+$ with the property that,
  for each $s\in\X$,
  the sequence $\{f'(n,s)\}_{n\in\N^+}$ of operators is
  a subsequence of the sequence $\{f(k,n,s)\}_{n\in\N^+}$.
  Thus $f(k,n,s)$ converges strongly to $1/2R(s)+1/2^{s+1}I$ as $n\to\infty$.
  This completes the proof.
\end{proof}
\medskip

Based on the above lemmas,
we can give the proof of Theorem \ref{existence-of-universal-semi-POVM}
as follows.

\begin{proof}[PROOF of Theorem \ref{existence-of-universal-semi-POVM}.]
  Let $\{e_i\}$ be a computable orthonormal basis for
  $\langle X,\fmi\rangle$.
  Let $f$ and $g$ be the functions given by Lemma \ref{universal-generator}
  and, for each $l\in\N^+$,
  let $R_l$ be a lower-computable semi-POVM on $\X$ such that,
  for each $s\in\X$,
  $f(l,n,s)$ converges strongly to $R_l(s)$ as $n\to\infty$.
  We first define an $f_M\colon\N^+\times\X\to\PO$ and
  a total recursive function $g_M\colon\N^+\times\X\to\N^+$ by
  \begin{eqnarray*}
    f_M(n,s)&=&\sum_{l=1}^n\frac{1}{2^l}f(l,n,s), \\
    g_M(n,s)&=&\max\{g(l,n,s)\mid 1\le l\le n\}.
  \end{eqnarray*}
  Obviously,
  the mapping
  $\N^+\times\X\times\N^+\times\N^+\ni (n,s,i,j)
  \longmapsto\ip{f_M(n,s)e_i}{e_j}$
  is a total recursive function and,
  for all $n$ and $s$,
  $f_M(n,s)$ is a $g_M(n,s)$-square rational operator on $\{e_i\}$.
  We also see that
  $f_M(n,s)\leqslant f_M(n,s)+f(n+1,n+1,s)\leqslant f_M(n+1,s)$.
  Since $f(l,n,s)\leqslant R_l(s)\leqslant I$,
  we have $f_M(n,s)\leqslant (1-2^{-n})I\leqslant I$.
  Thus, by Lemma \ref{bounded monotonic sequence},
  there exists an $M\colon\X\to\PO$ such that,
  for each $s\in\X$,
  $f_M(n,s)$ converges strongly to $M(s)$ as $n\to\infty$.
  We show that this $M$ is a universal semi-POVM.
  
  To begin with,
  we note that, for any $n,m\in\N^+$, any $s\in\X$, and any $x\in X$,
  \begin{eqnarray*}
    &&\Bigl\|\Bigl(\sum_{l=1}^n\frac{1}{2^l}R_l(s)\Bigr)x-M(s)x\Bigr\| \\
    &&\le \Bigl\|\sum_{l=1}^n\frac{1}{2^l}R_l(s)x-
    \sum_{l=1}^n\frac{1}{2^l}f(l,n+m,s)x\Bigr\|+
    \Bigl\|\sum_{l=n+1}^{n+m}\frac{1}{2^l}f(l,n+m,s)x\Bigr\| \\
    &&\hspace*{82mm}+\left\|f_M(n+m,s)x-M(s)x\right\| \\
    &&\le \sum_{l=1}^n\frac{1}{2^l}\left\|R_l(s)x-f(l,n+m,s)x\right\|+
    2^{-n}\|x\|+\left\|f_M(n+m,s)x-M(s)x\right\|.
  \end{eqnarray*}
  Here we use
  $\|f(l,n+m,s)x\|\le\|f(l,n+m,s)\|\|x\|\le \|x\|$.
  Thus, by choosing any one sufficiently large $m$
  for each sufficiently large $n$,
  we see that, for each $s\in\X$,
  $\sum_{l=1}^n 1/2^lR_l(s)$ converges strongly to $M(s)$
  as $n\to\infty$.
  For each $m\in\N^+$,
  since
  $\sum_{l=1}^n (1/2^l\sum_{s=1}^m R_l(s))\leqslant
  \sum_{l=1}^n 1/2^lI\leqslant I$ and
  $\sum_{l=1}^n (1/2^l\sum_{s=1}^m R_l(s))$ converges strongly to
  $\sum_{s=1}^mM(s)$,
  we have $\sum_{s=1}^mM(s)$ $\leqslant I$.
  It follows from $0\leqslant M(s)$ and Lemma \ref{bounded monotonic sequence}
  that $\sum_{s=1}^mM(s)$ converges strongly to
  $\sum_{s\in\X} M(s)\in\HO$ as $m\to\infty$ and
  $0\leqslant \sum_{s\in\X} M(s)\leqslant I$.
  Thus, since $f_M(n,s)-2^{-n}I\leqslant f_M(n+1,s)-2^{-n-1}I$,
  $M$ is a lower-computable semi-POVM on $\X$.
  
  Now, let $R$ be any lower-computable semi-POVM on $\X$.
  Then, by Lemma \ref{universal-generator},
  there is a $k$ with $1/2R(s)+(1/2)^{s+1}I=R_k(s)$.
  Since $1/2^kR_k(s)\leqslant\sum_{l=1}^\infty 1/2^lR_l(s)=M(s)$,
  we have $1/2^{k+1}R(s) \leqslant 1/2^{k+1}(R(s)+2^{-s}I)\leqslant M(s)$.
  Hence, $M$ is a universal semi-POVM.
\end{proof}

In the previous work \cite{T03},
we developed the theory of a universal semi-POVM
for a finite dimensional quantum system,
and we showed that, for every universal probability $m$,
the mapping $\X\ni s\longmapsto m(s)E$ is a universal semi-POVM
on a finite dimensional quantum system,
where $E$ is the identity matrix.
On the other hand,
as shown in the following proposition,
the corresponding statement does not hold for the infinite dimensional setting
on which we work at present.

\begin{proposition}\label{difference}
  Let $m$ be a universal probability.
  Then the mapping $\X\ni s\longmapsto m(s)I$ is not a universal semi-POVM.
\end{proposition}

\begin{proof}
  Let $\{e_i\}$ be an orthonormal basis for $X$, and
  let $P\colon\X\to\PO$ with $(P(s))(e_i)=\delta_{si}e_i$.
  Then $P$ is shown to be a POVM on $\X$.
  By Axiom \ref{axiom1} we see that $P$ is computable.
  Since $\|P(s)\|_2=1$ for every $s\in\X$,
  $P(s)$ is Hilbert-Schmidt for every $s\in\X$ and
  $\{\|P(s)\|_2\}_{s\in\X}$ is a computable sequence of real numbers.
  It follows from Theorem \ref{computable-lower-computable} that
  $P$ is a lower-computable semi-POVM on $\X$.
  
  Now, let us assume contrarily that
  the mapping $\X\ni s\longmapsto m(s)I$ is a universal semi-POVM.
  Then there exists a $c>0$ such that, for all $s\in\X$,
  $cP(s)\leqslant m(s)I$.
  Since $\ip{(P(s))e_s}{e_s}=1$,
  we have $c\le m(s)$ for all $s\in\X$.
  However, this contradicts the condition that $\sum_{s\in \X}m(s)\le 1$,
  and the proof is completed.
\end{proof}

Thus,
there is an essential difference between
finite dimensional quantum systems and infinite dimensional quantum systems
with respect to the properties of a universal semi-POVM.

\section{Extension of Chaitin's $\Omega$}
\label{omega}

In this section,
we introduce an extension of Chaitin's $\Omega$
as a partial sum of the POVM elements of a POVM measurement performed upon
an infinite dimensional quantum system.
Before that,
we give a relation between a universal semi-POVM and a universal probability.
We first show a relation
between a universal semi-POVM and a lower-computable semi-measure
in Proposition \ref{universal-probability-semi-POVM}.

\begin{proposition}\label{universal-probability-semi-POVM}
  Let $r$ be a lower-computable semi-measure,
  and let $M$ be a universal semi-POVM.
  Then there exists a $c>0$ such that, for all $s\in\X$,
  \begin{enumerate}
    \item $cr(s)I\leqslant M(s)$, and
    \item for all $x\in X$ with $\|x\|=1$,
      $cr(s)\le\ip{M(s)x}{x}$.
  \end{enumerate}
\end{proposition}

\begin{proof}
  The condition (ii) follows immediately from (i).
  Thus we show the condition (i).
  Since $r$ is a lower-computable semi-measure,
  $\sum_{s\in \X}r(s)\le 1$ and
  there exists a total recursive function
  $f'\colon\N^+\times \X\to\Q$
  such that, for each $s\in \X$,
  $\lim_{n\to\infty} f'(n,s)=r(s)$ and
  $\forall\,n\in\N^+\;\>0\le f'(n,s)\le f'(n+1,s)$.
  Let $\{e_i\}$ be a computable orthonormal basis for $X$ and,
  for each $n\in\N^+$,
  let $I(n)$ be the operator in $\HO$ such that
  $I(n)e_i=e_i$ if $i\le n$ and $I(n)e_i=0$ otherwise.
  We define $f\colon\N^+\times\X\to\HO$
  by $f(n,s)=f'(n,s)I(n)$. Since $0\leqslant I(n)\leqslant I(n+1)$,
  we have $0\leqslant f(n,s)\leqslant f(n+1,s)$.
  Since $I(n)$ converges strongly to $I$,
  $f(n,s)$ converges strongly to $r(s)I$ as $n\to\infty$.
  Obviously,
  $f(n,s)$ is an $n$-square rational operator on $\{e_i\}$, and
  the mapping
  $\N^+\times\X\times\N^+\times\N^+\ni(n,s,i,j)\longmapsto\ip{f(n,s)e_i}{e_j}$
  is a total recursive function.
  It follows from $\sum_{s\in\X}\{r(s)I\}\leqslant I$ that
  the mapping $\X\ni s\longmapsto r(s)I$ is
  a lower-computable semi-POVM on $\X$.
  Thus, from the definition of a universal semi-POVM,
  the condition (i) follows.
\end{proof}

Based on the above proposition, we can show the following.

\begin{theorem}\label{usP-implies-up}
  Let $M$ be a universal semi-POVM,
  and let $x\in X$ be computable with $\|x\|=1$.
  Then the mapping $\X\ni s\longmapsto \ip{M(s)x}{x}$ is
  a universal probability.
\end{theorem}

\begin{proof}
  Let $\{e_i\}$ be a computable orthonormal basis for $X$.
  We first define $c_i=\ip{x}{e_i}$.
  Then $\{c_i\}$ is a computable sequence of complex numbers
  which satisfies $x=\sum_{i=1}^\infty c_i e_j$.
  Since $M$ is a lower computable semi-POVM on $\X$,
  there exist an $f\colon\N^+\times\X\to\PO$ and
  a total recursive function $g\colon\N^+\times\X\to\N^+$ which satisfy
  the conditions (i), (ii), (iii), and (iv)
  in Definition \ref{def-pre-lower-computable-semi-povm}.
  Since $f(n,s)-2^{-n}I\leqslant M(s)$,
  we have $\ip{f(n,s)x}{x}-2^{-n}\le\ip{M(s)x}{x}$.
  It follows from
  $\ip{f(n,s)x}{x}=\sum_{i,j=1}^{g(n,s)} c_i\overline{c_j}\ip{f(n,s)e_i}{e_j}$
  that $\{\ip{f(n,s)x}{x}\}$ is a computable sequence of real numbers.
  Therefore,
  since $\lim_{n\to\infty} \ip{f(n,s)x}{x}-2^{-n}=\ip{M(s)x}{x}$,
  there exists a total recursive function $f'\colon\N^+\times \X\to\Q$
  such that,
  for each $s\in \X$,
  $\lim_{n\to\infty} f'(n,s)=\ip{M(s)x}{x}$ and
  $\forall\,n\in\N^+\;f'(n,s)\le f'(n+1,s)$.
  We then define a total recursive function
  $h\colon\N^+\times \X\to\Q$ by $h(n,s)=\max\{f'(n,s),0\}$.
  Since $\ip{M(s)x}{x}\ge 0$,
  we have $\lim_{n\to\infty} h(n,s)=\ip{M(s)x}{x}$ and
  $\forall\,n\in\N^+\;0\le h(n,s)\le h(n+1,s)$.
  We also have $\sum_{s\in\X}\ip{M(s)x}{x}\le \ip{Ix}{x}\le 1$.
  Thus the mapping $\X\ni s\longmapsto\ip{M(s)x}{x}$ is
  a lower-computable semi-measure.
  Finally, by Proposition \ref{universal-probability-semi-POVM},
  the theorem is obtained.
\end{proof}

Since any universal probability is not computable,
by Theorem \ref{usP-implies-up} we can show that
any universal semi-POVM is not a computable semi-POVM on $\X$.

Now, based on the intuition obtained from Theorem \ref{omega-equiv-universal},
we propose to define an extension $\hat{\Omega}$ of Chaitin's $\Omega$
as follows.

\begin{definition}[extension of Chaitin's $\Omega$ to operator]
  For each universal semi-POVM $M$,
  $\hat{\Omega}_M$ is defined by
  \begin{equation*}
    \hat{\Omega}_M\equiv\sum_{s\in\X} M(s).
  \end{equation*}
\end{definition}

Let $M$ be a universal semi-POVM.
Then, obviously, $\hat{\Omega}_M\in\PO$ and $\hat{\Omega}_M\leqslant I$.
We can further show that $cI\leqslant \hat{\Omega}_M$
for some real number $c>0$.
This is because, by Proposition \ref{universal-probability-semi-POVM},
there is a real number $c>0$ with the property that
$c2^{-s}I\leqslant M(s)$ for all $s\in\X$.

The following theorem supports the above proposal.

\begin{theorem}\label{prooerty-of-operator-omega}
  Let $M$ be a universal semi-POVM.
  If $x$ is a computable point in $X$ with $\|x\|=1$, then
  \begin{enumerate}
    \item there exists an optimal computer $V$ such that
      $\ip{\hat{\Omega}_Mx}{x}=\Omega_V$, and
    \item $\ip{\hat{\Omega}_Mx}{x}$ is a random real number.
  \end{enumerate}
\end{theorem}

\begin{proof}
  Since $\ip{\hat{\Omega}_Mx}{x}=\sum_{s\in\X}{\ip{M(s)x}{x}}$,
  by Theorem \ref{usP-implies-up} and Theorem \ref{omega-equiv-universal},
  Theorem \ref{prooerty-of-operator-omega} (i) follows.
  Since $\Omega_W$ is random for any optimal computer $W$,
  Theorem \ref{prooerty-of-operator-omega} (ii) follows.
\end{proof}

Let $M$ be any universal semi-POVM,
and let $x$ be any point in $X$ with $\|x\|=1$.
Consider the POVM measurement $\mathcal{M}$ described by the $M$.
This measurement produces one of countably many outcomes;
elements in $\X$ and one more something
which corresponds to the POVM element $I-\Omega_M$.
If the measurement $\mathcal{M}$ is performed
upon the state described by the $x$ immediately before the measurement,
then the probability that a result $s\in\X$ occurs is given by
$\ip{M(s)x}{x}$.
Therefore $\ip{\hat{\Omega}_Mx}{x}$ is the probability of getting
some finite binary string as a measurement outcome in $\mathcal{M}$.

Now, assume that $x$ is computable.
Recall that, for any optimal computer $V$,
$\Omega_V$ is the probability that
$V$ halts and outputs some finite string,
which results from infinitely repeated tosses of a fair coin.
Thus, by Theorem \ref{prooerty-of-operator-omega},
$\ip{\hat{\Omega}_Mx}{x}$ has the meaning of classical probability that
a universal self-delimiting Turing machine generates some finite string.
Hence $\ip{\hat{\Omega}_Mx}{x}$ has the meaning of probability
of producing some finite string
in the contexts of both quantum mechanics and algorithmic information theory.
Thus, in the case where $x$ is computable,
algorithmic information theory is consistent with quantum mechanics
in a certain sense.
Note further that,
even if $x$ is not computable,
quantum mechanics still insists that
$\ip{\hat{\Omega}_Mx}{x}$ has a meaning as probability, i.e.,
the probability of getting some finite binary string
in the measurement $\mathcal{M}$.

\section{Operator-valued algorithmic information theory}
\label{ovait}

We choose any one universal semi-POVM $M$
as the standard one for use throughout the rest of this paper.
The equation \eqref{eq: K_m}
suggests defining an \textit{operator-valued information content} $\OH(s)$
of $s\in\X$ by
\begin{equation}\label{ovK}
  \OH(s)\equiv -\log_2 M(s).
\end{equation}
Here $\log_2 M(s)$ is defined
based on the notion of \textit{continuous functional calculus}
(for the detail, see e.g.~the section VII.1 of \cite{RS80}).
We here note the following properties for this notion.

\begin{proposition}\label{operator-monotone}
  Let $S,T\in\HO$.
  Suppose that $aI\leqslant S$ for some real number $a>0$.
  Then $\log_2 S\in\HO$ and the following hold.
  \begin{enumerate}
    \item $\log_2(cS)=\log_2 S+(\log_2 c)I$ for any real number $c>0$.
    \item If $S\leqslant T$ then $\log_2 S\leqslant \log_2 T$.
  \end{enumerate}
\end{proposition}

Proposition \ref{operator-monotone} follows
the definition of the continuous functional calculus
(especially,
the proof of Proposition \ref{operator-monotone} (ii)
is given at e.g.~Chapter 5 of \cite{HY95}).
Since there is a real number $c>0$ with the property that
$c2^{-s}I\leqslant M(s)$ for all $s\in\X$,
by Proposition \ref{operator-monotone}
we see that $\OH(s)\in\HO$ for all $s\in\X$.
The above definition of $\OH(s)$ is also supported
by the following Proposition \ref{invariant-theorem}.
Let $S$ be any set,
and let $f\colon S\to\HO$ and $g\colon S\to\HO$.
Then we write $f(x)=g(x)+O(1)$
if there is a real number $c>0$ such that, for all $x\in S$,
$\|f(x)-g(x)\|\le c$,
which is equivalent to $-cI\leqslant f(x)-g(x) \leqslant cI$.

\begin{proposition}\label{invariant-theorem}
  Let $M$ and $M'$ be universal semi-POVMs.
  Then $\log_2 M(s)=\log_2 M'(s)+O(1)$.
\end{proposition}

\begin{proof}
  This follows immediately from Proposition \ref{operator-monotone}.
\end{proof}

By this proposition,
the equation \eqref{ovK} is independent of
the choice of a universal semi-POVM $M$ up to an additive constant.
We show relations between $\OH(s)$ and $\K(s)$
in the following theorem.

\begin{theorem}\label{operatorH-numberH}
  Let $x\in X$ with $\|x\|=1$.
  \begin{enumerate}
    \item There exists a real number $c>0$ such that
      $\ip{\OH(s)x}{x}\le \K(s)+c$ for all $s\in\X$.
    \item If $x$ is computable then $\ip{\OH(s)x}{x}=\K(s)+O(1)$.
  \end{enumerate}
\end{theorem}

\begin{proof}
  Since $2^{-\K(s)}$ is a lower-computable semi-measure,
  it follows from Proposition \ref{universal-probability-semi-POVM} that
  there is a $d>0$ with the property that
  $d2^{-\K(s)}I\leqslant M(s)$ for all $s\in\X$.
  By Proposition \ref{operator-monotone} (i) and
  the equality $\log_2 I=0$,
  we see that $\log_2(d2^{-\K(s)}I)=(-\K(s)+\log_2d)I$.
  Hence, by Proposition \ref{operator-monotone} (ii),
  we have $\OH(s)\leqslant(\K(s)-\log_2d)I$
  and therefore Theorem \ref{operatorH-numberH} (i) follows.
  
  Using the concavity of the real function $\log_2t$
  and the spectral decomposition of the Hermitian operator $\log_2 M(s)$,
  we can show that $\log_2\ip{M(s)x}{x}\ge\ip{(\log_2M(s))x}{x}$.
  In the case where $x$ is computable, by Theorem \ref{usP-implies-up},
  the mapping $\X\ni s\longmapsto\ip{M(s)x}{x}$ is
  a lower-computable semi-measure.
  By Theorem \ref{eup},
  there is a $c'>0$ such that $c'\ip{M(s)x}{x}\le 2^{-\K(s)}$ for all $s\in\X$.
  Hence Theorem \ref{operatorH-numberH} (ii) follows.
\end{proof}

In \cite{C87a} Chaitin developed a version of algorithmic information theory
where the notion of program-size is not used.
That is,
in the work he, in essence, defined $\K(s)$ as $-\log_2 m(s)$ 
for a universal probability $m$,
and showed several information-theoretic relations on $\K(s)$.
Thus we can develop the information-theoretic feature of
algorithmic information theory to a certain extent
even if we do not refer to the concept of program-size.
On the lines of this Chaitin's approach,
we show
in the following
that an information-theoretic feature can be developed
based on $\OH(s)$.
We first need the following theorem.

\begin{theorem}\label{usP-partial-recursive-func}
  Let $\psi\colon\X\to\X$ be a partial recursive function.
  Then the following hold.
  \begin{enumerate}
    \item There exists a real number $c>0$ such that, for all $s\in\X$,
      if $\psi(s)$ is defined then $cM(s)\leqslant M(\psi(s))$.
    \item There exists a real number $c>0$ such that, for all $s\in\X$,
      if $\psi(s)$ is defined then $\OH(\psi(s))\leqslant \OH(s)+cI$.
  \end{enumerate}
\end{theorem}

\begin{proof}
  Let $\{e_i\}$ be a computable orthonormal basis for $X$.
  Since $M$ is a universal semi-POVM,
  there exist an $f\colon\N^+\times\X\to\PO$ and
  a total recursive function $g\colon\N^+\times\X\to\N^+$ which satisfy
  the conditions (ii), (iii), and (iv)
  in Definition \ref{def-pre-lower-computable-semi-povm},
  and the condition that for each $s\in\X$,
  $f(n,s)$ converges strongly to $M(s)$ as $n\to\infty$.
  We can define $R\colon\X\to\BO$ by $R(s)=\sum_{\psi(t)=s}M(t)$,
  where the series converges strongly if $\psi^{-1}(s)$ is an infinite set.
  This limit exists
  by Lemma \ref{bounded monotonic sequence},
  since $\sum_{s=1}^l M(s)\leqslant I$ for any $l\in\N^+$ and
  $0\leqslant M(s)$.
  In the case of $\psi^{-1}(s)=\emptyset$,
  we interpret $\sum_{\psi(t)=s}M(t)$ as $0$.
  Obviously $0\leqslant R(s)$ for any $s\in\X$.
  Since $\ip{R(s)x}{x}=\sum_{\psi(t)=s}\ip{M(t)x}{x}$
  and $\sum_{s\in\X}\ip{M(s)x}{x}\le 1$,
  we see that $\sum_{s=1}^l R(s)\leqslant I$ for any $l\in\N^+$ and therefore,
  by Lemma \ref{bounded monotonic sequence}, $R$ is a semi-POVM on $\X$.
  
  Now, we enumerate
  the domain of definition of $\psi$.
  Let $t(k,s)$ be the $k$-th element in $\psi^{-1}(s)$ generated
  in the enumeration,
  and let $h(n,s)$ be the number of elements in $\psi^{-1}(s)$
  which are generated
  until the time step $n$ in the enumeration
  (possibly $h(n,s)=0$).
  We define $f'\colon\N^+\times\X\to\PO$
  by $f'(n,s)=\sum_{k=1}^{h(n,s)} f(n+k,t(k,s))$.
  It is then shown that $f'(n,s)-2^{-n}I\leqslant f'(n+1,s)-2^{-n-1}I$.
  We also define the total recursive function $g'\colon\N^+\times\X\to\N^+$ by
  $g'(n,s)=\max\{g(n+k,t(k,s))\mid 1\le k\le h(n,s)\}$.
  Then $f'(n,s)$ is a $g'(n,s)$-square rational operator on $\{e_i\}$.
  Since $f(n,s)-2^{-n}I\leqslant M(s)$,
  we have $f'(n,s)-2^{-n}I\leqslant R(s)-2^{-n-h(n,s)}\leqslant I$.
  Thus, by Lemma \ref{bounded monotonic sequence} again,
  for each $s\in\X$, there exists an $R'(s)\in\HO$ such that
  $f'(n,s)-2^{-n}I$ converges strongly to $R'(s)$ as $n\to\infty$.
  We show that $R(s)=R'(s)$ for all $s$.
  Since $f(n,s)-2^{-n}I\leqslant M(s)$,
  we have $|\ip{M(s)x}{x}-\ip{f(n,s)x}{x}|\le\ip{M(s)x}{x}+2^{-n}\|x\|^2$.
  Hence we see that if $1\le l<h(n,s)$ then
  \begin{eqnarray*}
    &&|\ip{R(s)x}{x}-\ip{f'(n,s)x}{x}| \\
    &&\le |\ip{R(s)x}{x}-
    \sum_{k=1}^{h(n,s)}\ip{M(t(k,s))x}{x}| \\
    &&\hspace*{40mm}
    +\sum_{k=1}^{h(n,s)}|\ip{M(t(k,s))x}{x}-\ip{f(n+k,t(k,s))x}{x}| \\
    &&\le |\ip{R(s)x}{x}-\ip{(\sum_{k=1}^{h(n,s)}M(t(k,s)))x}{x}| \\
    &&\hspace*{20mm}
    +\sum_{k=1}^l|\ip{M(t(k,s))x}{x}-\ip{f(n+k,t(k,s))x}{x}| \\
    &&\hspace*{40mm}
    +\sum_{k=l+1}^{h(n,s)}\ip{M(t(k,s))x}{x}+2^{-n}\|x\|^2.
  \end{eqnarray*}
  In the case where $\psi^{-1}(s)$ is an infinite set,
  since $\sum_{s\in\X}\ip{M(s)x}{x}<\infty$,
  by considering sufficiently large $n$
  for each sufficiently large $l$,
  we have $\lim_{n\to\infty}\ip{f'(n,s)x}{x}=\ip{R(s)x}{x}$.
  In the case where $\psi^{-1}(s)$ is a finite set,
  obviously the same holds.
  It follows that $\ip{R(s)x}{x}=\ip{R'(s)x}{x}$ for all $x\in X$ and $s\in\X$,
  and therefore $R(s)=R'(s)$ for all $s\in\X$.
  Hence $R$ is a lower computable semi-POVM on $\X$,
  and there is a real number $c>0$ such that,
  for all $s\in\X$, $cR(s)\leqslant M(s)$.
  From the definition of $R$, if $\psi(s)$ is defined
  then $M(s)\leqslant R(\psi(s))$.
  Thus Theorem \ref{usP-partial-recursive-func} (i) follows.
  By Proposition \ref{operator-monotone},
  we see that Theorem \ref{usP-partial-recursive-func} (ii) holds.
\end{proof}

We choose any one computable bijection $<s,t>$
from $(s,t)\in \X\times \X$ to $\X$.
Let $s, t\in \X$.
The \textit{joint information content} $\OH(s,t)$ of $s$ and $t$ is defined as
$\OH(s,t)\equiv \OH(<s,t>)$.
We then define
the \textit{conditional information content} $\OH(s|t)$ of $s$ given $t$
by the equation $\OH(s|t)\equiv \OH(t,s)-\OH(t)$.
Finally we define the \textit{mutual information content}
$\OH(s;t)$ of $s$ and $t$
by the equation $\OH(s;t)\equiv \OH(s)+\OH(t)-\OH(s,t)$.
Thus $\OH(s;t)=\OH(t)-\OH(t|s)$.
We can then show the following theorem
using Theorem \ref{usP-partial-recursive-func} (ii).
In particular, by Theorem \ref{opait} (i),
we have $\OH(s;t)=\OH(s)-\OH(s|t)+O(1)$.

\begin{theorem}\label{opait}\
  \begin{enumerate}
    \item $\OH(s,t)=\OH(t,s)+O(1)$ and $\OH(s;t)=\OH(t;s)+O(1)$.
    \item $\OH(s,s)=\OH(s)+O(1)$ and $\OH(s;s)=\OH(s)+O(1)$.
    \item $\OH(s,\lambda)=\OH(s)+O(1)$ and $\OH(s;\lambda)=O(1)$.
    \item $\exists\,c\in\R\;\>\forall\,s,t\in \X\;\>cI\leqslant \OH(s|t)$.
  \end{enumerate}
\end{theorem}

\begin{proof}
  Consider the total recursive function $\psi\colon\X\to\X$
  with $\psi(<s,t>)=<t,s>$.
  By Theorem \ref{usP-partial-recursive-func} (ii),
  there is a $c>0$ such that, for all $s,t\in\X$,
  $\OH(<t,s>)\leqslant \OH(<s,t>)+cI$.
  Thus Theorem \ref{opait} (i) follows.
  Next consider the function $\psi$ with $\psi(<s,s>)=s$.
  By Theorem \ref{usP-partial-recursive-func} (ii),
  there is a $c>0$ such that, for all $s\in\X$,
  $\OH(s)\leqslant \OH(<s,s>)+cI$.
  On the other hand, by considering the function $\phi$ with $\phi(s)=<s,s>$,
  we see that there is a $c'>0$ such that, for all $s\in\X$,
  $\OH(<s,s>)\leqslant \OH(s)+c'I$.
  Thus Theorem \ref{opait} (ii) follows.
  Similarly,
  by considering the functions
  $\psi$ with $\psi(<s,\lambda>)=s$ and $\phi$ with $\phi(s)=<s,\lambda>$,
  we have Theorem \ref{opait} (iii).
  Finally, by considering the function $\psi$ with $\psi(<s,t>)=s$,
  we have Theorem \ref{opait} (iv).
\end{proof}

The above relations can be compared with the following relations
in information theory except for the relation (v)
(see the discussion in Section \ref{discussion} for this exception).

\begin{theorem}\label{it}\
  \begin{enumerate}
    \item $H(X,Y)=H(Y,X)$ and $I(X;Y)=I(Y;X)$.
    \item $H(X,X)=H(X)$ and $I(X;X)=H(X)$.
    \item $H(X,Y)=H(X)$ and $I(X;Y)=0$
      if $Y$ takes any one fixed value with probability $1$,
      i.e., $H(Y)=0$.
    \item $0\le H(X|Y)$.
    \item $H(X,Y)\le H(X)+H(Y)$ and $0\le I(X;Y)$.
  \end{enumerate}
\end{theorem}

Here $X$ and $Y$ are discrete random variables,
and $H(X)$, $H(X,Y)$, $H(X|Y)$, and $I(X;Y)$ denote the
\textit{entropy},
\textit{joint entropy},
\textit{conditional entropy},
and \textit{mutual information},
respectively (see e.g.~\cite{CT91} for the detail of these quantities).
Thus, our theory built on $\OH(s)$ has
the formal properties of information theory to a certain extent.

\section{Discussion}
\label{discussion}

Based on a universal semi-POVM,
we have introduced $\hat{\Omega}_M$ which is
an extension of Chaitin's halting probability $\Omega_U$
to a measurement operator in an infinite dimensional quantum system,
and also we have introduced the operator $\OH(s)$
which is an extension of the program-size complexity $\K(s)$.
In algorithmic information theory, however,
$\Omega_U$ is originally defined through \eqref{Chaitin's omega}
based on the behavior of an optimal computer $U$,
i.e.,
$\Omega_U$ is defined as the probability that
the universal self-delimiting Turing machine which computes $U$ halts.
Likewise $\K(s)$ is originally defined as
the length of the shortest input for
a universal self-delimiting Turing machine
to output $s$.
Thus $\Omega_U$ and $H(s)$ are directly related to
a behavior of a computing machine.
Therefore,
in order to develop our operator version of
algorithmic information theory further,
it is necessary to find
more concrete definitions of $\hat{\Omega}_M$ and $\OH(s)$
which are immediately based on a behavior of some sort of computing machine.

In general,
a POVM measurement can be realized
by first interacting the quantum system on which
we make the POVM measurement with an ancilla system,
and then making a projective measurement upon
the composite system,
which consists of the original quantum system and the ancilla system.
This interaction is described by a unitary operator.
Let $U_M$ be such a unitary operator in
the POVM measurement described by an arbitrary universal semi-POVM $M$.
If we can identify a computing machine $\mathfrak{M}$ of some sort which
performs the unitary transformation $U_M$ in a natural way
in the POVM measurement,
then we might be able to give a machine interpretation to
$\hat{\Omega}_M$ and $\OH(s)$.
Note that the machine $\mathfrak{M}$ might be different kind of
computing machine from the so-called quantum Turing machine.
This is because
the unitary time evolution operator defined by a quantum Turing machine
makes local changes
on a quantum system, whereas $U_M$ makes global changes in general.
We leave the development of this line to a future study.

Now,
by defining $\K(s)$ as $-\log_2 m(s)$ for any one universal probability $m$,
\cite{C87a} proved
the following theorem,
which corresponds to the inequality in information theory
called \textit{subadditivity}, i.e., Theorem \ref{it} (v).

\begin{theorem}[subadditivity]\label{subadditivity}
  $\exists\,c\in\R\;\>\forall\,s,t\in \X\;\>c\le\K(s;t)$.
\end{theorem}

Here $\K(s;t)$ was defined as $\K(s)+\K(t)-\K(<s,t>)$ in \cite{C87a}.
Because of the non-commutativity of operators in $X$, however,
it is open to prove the corresponding formula for our $\OH(s;t)$.
In the proof of Theorem \ref{subadditivity} given in \cite{C87a},
the product $m(s)m(t)$ is considered.
In general, a product of two POVM elements has no physical meaning
unless they commute.
For a universal semi-POVM $M$,
it would seem difficult to prove the commutativity of
$M(s)$ and $M(t)$ for distinct $s$ and $t$.
Thus $M(s)M(t)$ seems to have no physical meaning
as a product of two POVM elements.
Hence the difficulty in proving the subadditivity for our $\OH(s;t)$
seems
to justify our interpretation of a universal semi-POVM
as measurement operators
which describe a quantum measurement performed upon a quantum system.
Note that, as is shown in \cite{T03},
we have the subadditivity in \textit{finite} dimensional setting.
This is because
$m(s)E$ is a universal semi-POVM in a finite dimensional linear space
for any universal probability $m$,
where $E$ is the identity matrix.
Obviously, $m(s)E$ and $m(t)E$ commute in this case.

\section*{Acknowledgments}

The author is grateful to
the 21st Century COE Program of Chuo University
for the financial support.


\end{document}